\documentclass[]{ccs-thesis}

\group{Telecommunication Networks (TKN)}

\author{Jorge Torres Gómez}
\orcid{0000-0001-9523-048X}
\birthday{}
\birthplace{}
\studentid{}

\title{Timely Information Updates in Molecular Communication: Age of Information in Nanoscale Systems}
\title{How much is time a critical factor in IoBNT networks?}
\title{How Time-Sensitive are IoBNT Networks?\\An Age of Information Perspective for In-Body Monitoring}

\thesistype{Habilitation's Thesis in Electrical Engineering}\thesiscite{Habilitation's Thesis}
\thesisinstitution{School of Electrical Engineering and Computer Science}
\thesisschool{TU Berlin}
\thesiscity{Berlin, Germany}

\advisors{Prof.\ Dr.-Ing.\ habil.\ Falko Dressler}
\referees{Prof.\ Ph.D.\ Massimiliano Pierobon, Prof.\ Ph.D. Tuna Tugcu}
\thesubmissiondate{4. March 2025}
\placeandyear{Berlin 2025}

\orcid{0000-0001-9523-048X}
\faculty{Fakultät IV --- Elektrotechnik und Informatik}
\university{Technischen Universität Berlin}
\refereechair{Prof.\ Dr.\ Committee Chair}
\thedefensedate{6. April 2025}

\acrodef{MC}{molecular communication}
\acrodef{IoBNT}{internet of bio-nano-things}
\acrodef{QS}{quorum sensing}

\acrodef{AoI}{age of information}
\acrodef{PAoI}{peak age of information}
\acrodef{BER}{bit error rate}
\acrodef{LTV}{linear time-variant}
\acrodef{CIR}{channel impulse response}
\acrodef{RMS}{root mean square}
\acrodef{SNR}{signal to noise ratio}
\acrodef{PDF}{probability density function}
\acrodef{BPSK}{binary phase shift keying}
\acrodef{AM}{amplitude modulation}
\acrodef{FIR}{finite impulse response}
\acrodef{PER}{packet error rate}
\acrodef{BVS}{BloodVoyagerS}
\acrodef{QoS}{quality of service}

\usepackage{lipsum}

\usepackage{booktabs}
\usepackage{rotating}
\usepackage{multirow}
\usepackage{bigstrut}

\usepackage[skip=0.33\baselineskip]{caption}

\usepackage[flushleft]{threeparttable}
\usepackage{xcolor}
\usepackage{colortbl} 
\usepackage{longtable}
\usepackage{adjustbox}
\usepackage{lscape} 

\usepackage{pdfpages}

\Crefformat{figure}{#2Fig.~#1#3}
\Crefformat{equation}{#2Eq.~(#1)#3}

\usepackage{nicefrac}

\usepackage{footnote}
\usepackage{tablefootnote}

\usepackage{tcolorbox}
\definecolor{myblue}{rgb}{0.651, 0.788, 0.925}

\Crefformat{figure}{#2Fig.~#1#3}
\Crefformat{equation}{#2Eq.~(#1)#3}
\Crefformat{equations}{#2Eqs.~(#1)#3}

\usepackage{hyperref}
\hypersetup{
    hidelinks
}

\begin{document}
\sloppy

\pagenumbering{roman}

\maketitle






        











\thispagestyle{empty}

\cleardoublepage

\begin{otherlanguage*}{ngerman}
{\Large\noindent%
Erklärung \newline}

\noindent%
Hiermit versichere ich, dass ich die vorliegende Arbeit eigenständig ohne Hilfe Dritter und ausschließlich unter Verwendung der aufgeführten Quellen und Hilfsmittel angefertigt habe.
Alle Stellen, die den benutzten Quellen und Hilfsmitteln unverändert oder sinngemäß entnommen sind, habe ich als solche kenntlich gemacht.

\noindent%
Sofern generische KI-Tools verwendet wurden, habe ich Produktnamen, Hersteller, die jeweils verwendete Softwareversion und die jeweiligen Einsatzzwecke (z.B. sprachliche Überprüfung und Verbesserung der Texte, systematische Recherche) benannt.
Ich verantworte die Auswahl, die Übernahme und sämtliche Ergebnisse des von mir verwendeten KI-generierten Outputs vollumfänglich selbst.

\noindent%
Ich erkläre weiterhin, dass ich die Arbeit in gleicher oder ähnlicher Form noch keiner anderen Prüfungsbehörde vorgelegt habe.

\noindent%
Die \href{https://www.static.tu.berlin/fileadmin/www/10000060/FSC/Promotion___Habilitation/Dokumente/Grundsaetze_gute_wissenschaftliche_Praxis_2017.pdf}{Satzung zur Sicherung guter wissenschaftlicher Praxis an der TU Berlin} vom 8. März 2017 habe ich zur Kenntnis genommen.

\end{otherlanguage*}

\vspace{2cm}

\noindent
\begin{tabular}{@{}l p{2cm} l@{}}
    Berlin, 14. Mai 2026 & & \rule{6cm}{0.5pt} \\
    Ort, Datum & & Jorge Torres Gómez \\
\end{tabular}

\chapter*{Acknowledgments}
\addcontentsline{toc}{chapter}{Acknowledgments}
\begin{otherlanguage*}{american}

I kindly express my gratitude to my fellow colleagues who helped me with the revision of this thesis.
They offered my valuable feedback, all included here, which made this content more readable.
Thanks to Agon Memedi, Lisa Y. Debus, Sigrid Dimce, Doğanalp Ergenç, Jorge Luis González, Karel Toledo, and Prof. Adam Wolisz for your time and the deep reading you made.
I also thank my supervisor Prof. Falko Dressler, for all the support over all these years, the academic comfort he provided in the TKN group, and the revision of this thesis.
The content of this cumulative thesis is also part of collaborative work with researchers who also dedicated efforts to joint publication: Anke Juestner, Joana Angjo, Jennifer Simonjan, Bige D. Unluturk, Ketki Pitke, Lukas Stratmann, and Josep Jornet.
Developments within this thesis have also been supported by NaBoCom, funded by the German Research Foundation (DFG) under grant DR 639/21-3, and the project IoBNT, funded by the Federal Ministry of Education and Research (BMBF, Germany) under grant 16KIS1986K.
\end{otherlanguage*}

\newpage

\thispagestyle{empty} 

\vspace*{25em}
\begin{flushright}
\textit{
To my grandparents, Gloria and Leonardo, whose steps as educators I have had the good fortune to follow.}
\end{flushright}

\newpage

\chapter*{Abstract}
\addcontentsline{toc}{chapter}{Abstract}
\begin{otherlanguage*}{american}

This thesis formulates a theoretical framework to evaluate the monitoring capability of \ac{IoBNT} networks.
We consider an scenario where \ac{IoBNT} network comprises nanosensors that passively flow in the bloodstream and can detect biomarkers related to potential diseases.
Nanosensors can also report this detection to external gateways placed on the skin's surface that host a monitor device.
In this way, the nanosensors fetch an artificial point-to-point communication channel between the disease region in the human body and the monitor device.
Similar to standard communication channels, some packets might get to the destination straight, while others can get lost (through vessel circuits other than the gateway). 
Following this network structure, we evaluate the network monitoring capability over this artificial channel using the \ac{AoI} concept.
The \ac{AoI} concept incorporates the joint integration of sample generation (at the disease region), carrying (nanosensor travel through the human vessels), and delivery (nanosensor-to-gateway) as random events.
These three random events are represented through (i) a Markov model, which follows the physiology of the cardiovascular system, and (ii) channel models of reported nanocommunication technologies.
Furthermore, we evaluate the transition probabilities for the Markov model using a cardiovascular system simulator, which consists of a low-complexity electric circuit model representing the human vessels.
As for the nanosensor-to-gateway link model, we model two well-known schemes with ultrasonic and terahertz channels.
Aiming to assess the monitoring capabilities of this network, we integrate these components within the \ac{AoI} framework and illustrate the most relevant figures for information freshness with the average \ac{PAoI} metric.
Under realistic physiological assumptions and communication models, we evaluate the order of magnitude in the tens of seconds to display fresh information on the monitor.
In this way, this network can monitor processes at the tissue level, such as bacteria infections, and more adequate network architectures are needed to monitor on a cellular scale, where processes occur in a timescale of less than tens of seconds.
\end{otherlanguage*}

\acresetall

\chapter*{Zusammenfassung}
\addcontentsline{toc}{chapter}{Kurzfassung}
\begin{otherlanguage*}{ngerman}

Diese Arbeit entwickelt einen theoretischen Rahmen zur Bewertung der Überwachungsfähigkeit von IoBNT-Netzwerken zur rechtzeitigen Erkennung und Behandlung potenzieller Krankheiten im menschlichen Körper. Unser Ziel ist ein IoBNT-Netzwerk aus Nanosensoren, die passiv im Blutkreislauf fließen, Biomarker erkennen und diese an externe Gateways auf der Hautoberfläche melden. Auf diese Weise stellen die Nanosensoren einen künstlichen Punkt-zu-Punkt-Kommunikationskanal zwischen der erkrankten Region und dem Überwachungsgerät her. Ähnlich wie bei Standardkommunikationskanälen können einige Pakete direkt an ihr Ziel gelangen, während andere über andere Gefäßkreisläufe verloren gehen können. Anhand dieser Netzwerkstruktur evaluieren wir die Überwachungsfähigkeit mithilfe des AoI-Konzepts, das Probengenerierung, Transport und Abgabe als Zufallsereignisse integriert. Diese Ereignisse werden durch ein Markov-Modell der Herz-Kreislauf-Physiologie sowie Kanalmodelle bekannter Nanokommunikationstechnologien dargestellt, wobei wir die Übergangswahrscheinlichkeiten mithilfe eines wenig komplexen Stromkreismodells bewerten. Für das Nanosensor-zum-Gateway-Verbindungsmodell modellieren wir zwei Verfahren mit Ultraschall- und Terahertz-Kanälen. Um die Überwachungsfähigkeiten zu bewerten, integrieren wir diese Komponenten in das AoI-Framework und veranschaulichen die Informationsaktualität mit der durchschnittlichen PAoI-Metrik. Unter realistischen physiologischen Annahmen ermitteln wir eine Größenordnung im Zehn-Sekunden-Bereich für aktuelle Informationen auf dem Monitor. Auf diese Weise kann dieses Netzwerk Prozesse auf Gewebeebene überwachen; für die zelluläre Ebene mit Zeitrahmen von weniger als zehn Sekunden sind jedoch geeignetere Netzwerkarchitekturen erforderlich.

\end{otherlanguage*}

\acresetall

\cleardoublepage
\tableofcontents

\cleardoublepage
\pagenumbering{arabic}

%

\acresetall

\part{Scientific Part}
\acresetall
\chapter[Introduction]{Introduction\\ \large{Relevance of the Age of Information concept for \acs{IoBNT}}}
\label{sec_introduction}

Medical treatments often require the timely monitoring of the disease evolution to assess the response to medications effectively.
Today's monitoring follows standard procedures, which require taking samples, such as blood, and processing them in medical laboratories.
The process takes hours; for instance, conventional methods require the culture of bacteria from samples to raise levels of detection, which last for $\SIrange{48}{72}{\hour}$~\cite{akyildiz2020panacea}.
The fastest lab strategies report turnaround times (the interval from sample collection to result reporting) within the hour \cite{lee2022strategies}.
Shifting to speedy monitoring procedures requires a different setup other than extracting and analyzing samples in a lab.

\begin{wrapfigure}{R}{0.5\textwidth}
\centering
\includegraphics[width=\linewidth]{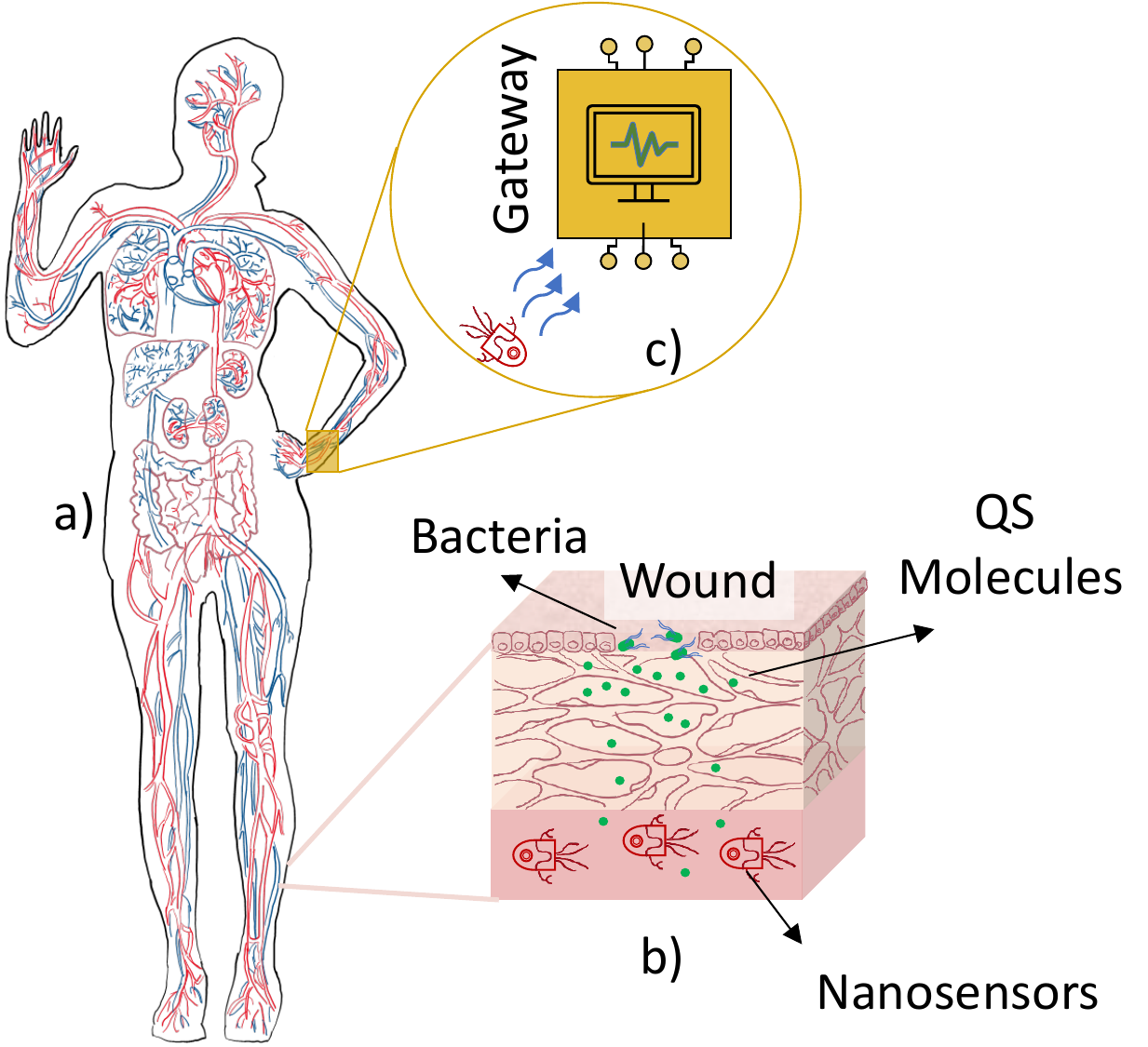} 
\caption{Concept of in-vivo disease detection scheme.}
\label{fig_architecture}
\end{wrapfigure}

Literature reports a solution to the above problem: Embedding the monitoring system within the body and performing wireless telemetry to external devices, as follows from the paradigm of connecting the cell to the internet in \cite{akyildiz2020panacea}.
Toward this conception, the \acf{IoBNT} network has received increased attention in the research community; see the survey references in \cite{akyildiz2020panacea,akyildiz2019moving,etemadi2023abnormality,kulakowski2020nano}.
The network encompasses nanosensors implanted within the human body with the capabilities of detecting disease-related biomarkers, see a representation in \Cref{fig_architecture} b), and gateways placed on the skin surface to connect the nanosensors to the Internet; see \Cref{fig_architecture} c).

Perhaps the reader will consider a description of a far-fetched vision at this point, but the technological development of bioelectronics components glimpses the potential development for telemetry within the human body.
Bioelectronics is already a mature field for gauging chemical compounds and conveying those in the electric domain, see recent literature review and examples in \cite{guo2024electrochemical, rivnay2025integrating,saleh2024bioelectronic}.
In-sensory computing also promises functional units in the nanoscale with the capabilities of joint sensing and processing as digital calculators; see \cite[Fig. 7]{nirmal2024advancements}.
Developments also refer to electromagnetic interfaces through the body tissues between implanted nanosensors and external gateways, as attached to the skin surface.
Analog front-ends have been modeled in the terahertz band to enable wireless communication links through the body tissues, as summarized in \cite{yang2020comprehensive}.
Signal generators as in \cite{dong2020recent} and antennas as in \cite{saeed2021body-centric, abadal2019graphene} of micro-scale size demonstrate means to enable intrabody communication links between nanosensors and gateways.

Holding the above description of technological advancements, the question remains regarding how time-sensitive \ac{IoBNT} networks may be in conveying information.
Within the scenario in \Cref{fig_architecture}, where nanosensors patrol the human body as flowing in the human vessels, can the wrist monitor display new information the nanosensor reports about the bacterial infection on the leg?
Bacteria duplicate their population number every $\SIrange{20}{25}{\min}$ at the tissue level, see \cite[page 279]{alberts2013essential}; which would require a monitoring update in the minute-time range.
Larger stringent times occur at the cell level, live processes as metabolic pathways occur much faster: the ribosomes in the cell take around $\SI{20}{\second}$ to several minutes to synthesize proteins (\cite[page 232]{alberts2013essential}) and the protein components (like the amino acids) are assembled in the units of milliseconds (\cite[page 229]{alberts2013essential}), which require for the monitor to display information on the same time scale.

\begin{tcolorbox}[colframe=myblue, colback=gray!10, sharp corners=southwest]
Assessing whether the \ac{IoBNT} network meets these time constraints requires a direct evaluation of time-related metrics, and we suggest an analytic framework for doing so.
This thesis examines explicitly how time-sensitive \ac{IoBNT} networks are and aims to unlock this question by evaluating information freshness.
\end{tcolorbox}
\noindent


Information freshness refers to the elapsed time between events displayed at the monitor and their generation time \cite{pappas2023age}.
The less time elapsed between the information displayed and the current development of the event, the better the monitor is updated about the events at the source.
The concept of information freshness is analytically formulated in the literature with the \ac{AoI} function; see \cite[Fig. 1]{yates2019age}.
This function displays the age of the received packet in the units of seconds\footnote{The age of a packet is the time elapsed since its generation.} at the monitor.
Metrics of this function, like the average or the peaks, provide discernment for information freshness.
For instance, considering the scheme in \Cref{fig_architecture}, a new packet is generated whenever the nanosensor passes by the infection location in b), and the packet starts to age with time when it is being received at the gateway in c), where it is displayed.
At the gateway monitor, the packet continues to age until the next packet arrives, at which point the monitor is updated.
Then, the age of the information displayed on the monitor follows that of the second packet.
In this way, the \ac{AoI} function is constructed and effectively accounts for the information freshness at the monitor.

\begin{tcolorbox}[colframe=myblue, colback=gray!10, sharp corners=southwest]
The relevance of the \ac{AoI} concept lies in yielding a metric for the monitor's update period.
In fact, the monitor's update period pertains to the timeliness of the information shown instead of its refresh rate.\footnote{The monitor can refresh the information at high speed, but the information can still be outdated.}
Besides, the \ac{AoI} jointly articulates the random events influencing the capability to display fresh information: generation, carrying, and delivery.
This formulation in the \ac{AoI} function enables us to examine the \ac{IoBNT} network resources, incorporating transmission capability, throughput, and delay.\footnote{From a design perspective, the concept allows for the comprehensive assessment of system parameters such as the number of nanosensors, lifespan, and the gateway's placement concerning information freshness.}
\end{tcolorbox}

Alongside the development of the \ac{AoI} concept in the \ac{IoBNT} network, we provide orders of magnitude to comparatively assess capabilities of monitoring biological processes in the human body.
In the following sections, we provide more context related to this concept within the literature, describe the system model, and formulate the strategy to evaluate the average \ac{PAoI} metric.



\section[Information Freshness in IoBNT]{Information Freshness in IoBNT\\ --A Literature Review}

The evaluation of information freshness is still in the early stage of development in \ac{IoBNT} networks.
Attention to this topic is explicitly stated in the literature as follows on a central premise of \ac{IoBNT} networks: shifting from long-lasting laboratory blood testing to \textit{in-situ} detection.
Health conditions are monitored through nanosensors within the human body to indicate harmful concentration levels before symptoms appear, and timely information is required at remote locations.
Examples of use cases are the monitoring by nanodevices of bacterial infections \cite[Sec. IV]{akyildiz2020panacea}, viral load, reaction to infections such as sepsis, and vessel obstruction due to atherosclerosis as summarized in \cite[Sec. II]{canovascarrasco2020understanding}) and \cite[Sec. II]{asorey-cacheda2022bridging}.
Following this trend, recently published surveys highlight real-time features of \ac{IoBNT} networks as envisioned design criteria \cite{kulakowski2020nano,etemadi2023abnormality,canovascarrasco2020understanding,akyildiz2020panacea}; nonetheless, the timely flow of information within the network is rarely formulated.

Most research in \ac{IoBNT} networks elaborate on their theoretical and experimental aspects, which are the bases to evaluate the information freshness.
\ac{IoBNT} is systematically studied as a functional platform integrating sensors and gateways \cite{etemadi2023abnormality,kulakowski2020nano}.
Specifically, interfaces between biological and electrical domains have been reported in synaptic links \cite{veletic2019synaptic}, optical links \cite{jornet2019optogenomic}, within the electrical domain \cite{kim2019redox}, and with electromagnetic signals as described in \cite{yang2020comprehensive}.
Theoretical developments of communication schemes that apply to the human body include the evaluation of the \ac{CIR} \cite{jamali2019channel}, channel parameter estimation~\cite{huang2022survey}, modulation and demodulation \cite{kuran2021survey,gursoy2022towards}, and coding \cite{hofmann2023coding}.

Lining towards examining the timely reporting of events, the work in \cite{canovascarrasco2020understanding} formulates a Markov chain of variable size, where each state refers to time slots for transmissions.
The Markov chain evaluates the stationary probabilities that the nanosensor performs a transmission, yielding the delay in communicating a detection.
Yet, this calculation is insufficient to describe the information freshness --delay is a necessary component but insufficient to maximize freshness.
The generation time needs to be integrated, and the interplay with delay needs to be analyzed.

The integration of generation, carrying, and delivery to assess information freshness can be aligned as introduced in the aforementioned papers E~\cite{torres-gomez2023age} and F~\cite{torres-gomez2022age}; which is the addressed topic in \Cref{sec_AoI}.
In the literature, the \ac{AoI} topic starts to gain some follow-up attention with a direct mention in research related to coding, as referenced in \cite{hofmann2023coding}, localization techniques as~\cite{lau2024inflammationbased} and within the overall description of \ac{IoBNT} networks as~\cite{cevallos2024molecular,jani2023bio-inspired}.
More straightforwardly, the work in ~\cite{pal2024age} developed a simulator for monitoring biomarkers and quantifying the average \ac{PAoI} metric in more realistic scenarios where nanosensors have a limited operational lifetime.
We expect further developments on the \ac{AoI} topic, as more insights are needed to design \ac{IoBNT} networks that promptly display and react to potential diseases in the human body; see also potential outlooks in \Cref{sec_future}. 

%

\section{System Model}

We develop the \ac{AoI} concept with the \ac{IoBNT} network illustrated in \Cref{fig_architecture}.
The network consists of nanosensors, here conceptualized to flow through the vessels passively as in \Cref{fig_architecture}~a) and a gateway device (placed in the skin's surface) to enable a communication link with the nanosensors, as represented in \Cref{fig_architecture}~b).
Within this setup, a network of nanosensors with the capabilities of monitoring in situ and communicating stands at the core of this network. 
As illustrated in the figure, the nanosensor will detect the infection at the location (in \Cref{fig_architecture}~b) and report the detection to a more powerful monitor device, as represented in \Cref{fig_architecture}~c).

This illustration depicts an artificial communication link within the human body.
Once the nanosensor flows through the infection location (in the ankle), it will detect \ac{QS} molecules released by the bacteria, and the results of this detection are carried with the nanosensor to the gateway location (in the wrist), where the information packet is delivered and displayed.
Within communication theory, the infection location is the source, nanosensors perform as information carriers, and human vessels are the channels that link the event in the infection location to the monitor, which performs as the destination node.
That is, the nanosensors flow with the bloodstream sets an artificial communication channel between the source and the destination.

%



To display an infection as a result of a wound in the leg, we must first model the following three components:

\begin{wrapfigure}{R}{0.45\linewidth}
	\centering
	\includegraphics[width=\linewidth]{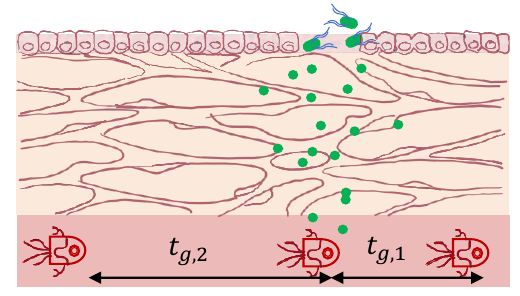}
	\caption{Visualizing the generation time period at the source. See instances with the variables $t_{g,1}$ and $t_{g,2}$.}
	\label{fig_generation_time_intro}
\end{wrapfigure}

\noindent
\textbf{1. The source:} The bloodstream drives the nanosensors, which travel randomly through the bifurcations within the human cardiovascular system and may not encounter the infection location.
For example, from the arcus aorta, next to the left ventricle in the heart, the bloodstream randomly guides the nanosensor to the lower body (through the thoracic and femoral aortas), where the infection is located or deviates it to the upper body (through the {superior vena cava}).
As such, the infection location is modeled as a source that randomly displays ones (whenever a nanosensor passes by the infection location) or zero otherwise.\footnote{For simplicity, we model a binary source. However, the source can represent the concentration level of a given biomarker, such as the number of \ac{QS} molecules released by bacteria.}
This source exhibits a random period of their pulses; as illustrated in \Cref{fig_generation_time_intro} with the instances $t_{g,1}$ and $t_{g,2}$, their generation time is dependent on the traveling distance among subsequent nanosensors.\footnote{For simplicity, here we assume that infection is detected whenever the nanosensor is within the vessel segment that corresponds to the infection location.}
    
\noindent
\textbf{2. The channel:}
Once the nanosensor detects the infection, it carries the information along the random path of the cardiovascular system to the gateway location.
Traveling from the infection location to the heart is always a deterministic path, as all the vessels sink in the heart's right ventricle~\cite{guyton2015guyton}; however, from the heart to the gateway location is again a random process.
The nanosensor might be randomly guided through the {brachialis aorta} on the left arm, where the gateway is located, or to another vessel circuit, thereby missing the opportunity to report the detection.
As such, with the communication theory terminology, this channel introduces a random delay.
The delay is minimal whenever the nanosensor travels straight from the infection location to the gateway, following the right vessels' bifurcations.
However, if the nanosensor loops in vessel circuits other than the gateway or the infection, the delay increases and might also reach infinite time.

\noindent
\textbf{3. The receiver:} Once the nanosensor is near the gateway, it transmits the message to the gateway device.
The message in this case is binary, with the one representing a successful detection and the zero a non-detection at the given infection location.
The transmission will be successful or not depending on the communication link quality between the nanosensor and the gateway.
The gateway will perform as the receiver and, as such, be modeled with a probability of successful reception.
%
%

The randomness of the three components above must be modeled using the physics of nanosensor mobility patterns within the cardiovascular system.
Then, using these models, we formulate the related \ac{AoI} metric, here developed with the average \ac{PAoI} \cite{yates2021age}.
Finally, with the average \ac{PAoI}, we appraise the time-sensitivity of the \ac{IoBNT} network.

%

\section{Scientific Approach}
\label{sec_strategy}

Our strategy follows from the abstraction of the above three components into the three random variables:
\begin{enumerate}
[label=\textbf{- Random Event~\arabic*:}, labelindent=0pt, wide, labelwidth=!]
\item The nanosensor meets the infection location.
\item The nanosensor travels from the infection to the gateway location.
\item The nanosensor successfully delivers the information to the gateway device.
\end{enumerate}

The fulfillment of these three events will ultimately define the monitoring capabilities of the \ac{IoBNT} network, and their joint evaluation is directly conveyed within the \ac{AoI} concept.
Formulating the \ac{AoI} concept realizes the following abstractions:
\begin{itemize}
    \item The source is abstracted within the generation time, here denoted as $T_g$  and related to the Random Event~1.
    This random variable will depend on how frequently a nanosensor visits the infection location, as illustrated in \Cref{fig_architecture}~b).
    \item The information delivery is abstracted within the system time -related to Random Events~2 and~3.
    This random variable depends on traveling time along human vessels in \Cref{fig_architecture}~b), here denoted as $T_d$, and the successful delivery of transmissions (ocurring in \Cref{fig_architecture}~b), here evaluated as $(1-p_\mathrm{loss})$, where $p_\mathrm{loss}$ is the probability of failure delivery.
    \item The concept of \ac{AoI} appraises the impact of these two variables mainly with two metrics: the average \ac{AoI} and the average \ac{PAoI}; see the definition in \cite[Def. 2.1.2 and Sec. 3.3]{kosta2017age}.
    These two metrics assess the average time the monitor refreshes the source status.
    In this thesis, we develop the average \ac{PAoI} metric due to its mathematical tractability compared to the average \ac{AoI} metric.
\end{itemize}


With the average \ac{PAoI} metric, we trace the following strategy as the main leading steps in this thesis:
%


%
\begin{enumerate}
    \item Formulate a closed-form expression for the average \ac{PAoI} metric.
    This formulation is developed in \Cref{sec_average_PAoI} with \Cref{eq_PAoI}, here reproduced as
    \begin{equation}
        \Delta^{(p)}=\frac{1}{1-p_\mathrm{loss}}\mathbf{E}[T_g]+\mathbf{E}[T_d],
    \end{equation}
    where $T_g$ is the generation time (related to Random Event~1), $T_d$ is the random delay (related to Random Event~2), $p_\mathrm{loss}$ packet loss probability (related to Random Event~3).
    From there, we realize the need to characterize the variables $T_g$, $T_d$, and $p_\mathrm{loss}$, and $\mathbf{E}[T_g]$ is the expectation operator.
    \item The $p_\mathrm{loss}$ value is characterized with the communication technology.
    Here, we develop this calculation for ultrasonic and terahertz emissions; see a description in \Cref{sec_comm}.
    \item The variable $T_g$ is modeled with the exponential distribution of parameter $\lambda$; see the rationale in \Cref{sec_average_PAoI}.
    The $\lambda$ parameters refer to the concentration level of the nanosensor in the infection location and are evaluated with a Markov model.
    %
    \item The expected value for the random variable $T_d$ is also evaluated with the Markov model.
    A closed-form expression is derived in \Cref{eq_delay_2}, looking at the cardiovascular system as a connection of closed-loop circuits; see the rationale in \Cref{sec_average_PAoI}.
    \item The human cardiovascular system is abstracted with a Markov model.
    States in the Markov model refer to vessel segments and their transition probabilities to the dynamic in the bifurcation of the human vessels.
    The stationary probabilities of the Markov model allow for the evaluation of $T_g$ and their transition probabilities for the variable $T_d$.
    \item We develop a simulator for the human cardiovascular system in the electric domain.\footnote{The design for the electric circuit representation is accessible in \url{https://github.com/tkn-tub/frankenstein}}
    This simulator evaluates the transition probabilities for the Markov model with the blood flow ratio of the vessels' bifurcations; see the rationale in \Cref{sec_frankesntein}.
\end{enumerate}
Fulfilling these steps in bottom-to-top order (sequential sort), we evaluate the average \ac{PAoI} metric along this thesis.

%

\section{Thesis Organization}

In this thesis, we formulate analytic expressions for the three random variables $T_g$, $T_d$, and $p_\mathrm{loss}$, as well as the average \ac{PAoI} with the following structure in chapters:

\begin{enumerate}[label=\textbf{Chapter~\arabic*:}, labelindent=0pt, wide, labelwidth=!]
\setcounter{enumi}{1} 
    \item This chapter introduces the modeling of the random mobility of nanosensors within the human vessels.
    This chapter summarizes the Markov model to evaluate the path trajectory of nanosensors within the blood vessels.
    The content of this chapter related to the contributions\\    
    \begin{itemize}
        \item \textbf{Paper A:} Jorge Torres Gómez, Anke Kuestner, Jennifer Simonjan, Bige Deniz Unluturk and Falko Dressler, "Nanosensor Location Estimation in the Human Circulatory System using Machine Learning," IEEE Transactions on Nanotechnology, vol. 21, pp. 663–673, October 2022.
        \item \textbf{Paper B:} Jorge Torres Gómez, Jorge Luis González Rios and Falko Dressler, "Electric Circuit Representation of the Human Circulatory System to Estimate the Position of Nanosensors in Vessels," Elsevier Nano Communication Networks, vol. 40, pp. 100499, July 2024.
    \end{itemize}
    \item This chapter introduces the model of successfully delivering information to the gateway.
    Here, we evaluate the communication link between the nanosensors and the gateway through a communication link in the electromagnetic and ultrasonic domains.
    This chapter is related to the contributions:
    \begin{itemize}
        \item \textbf{Paper C:} Jorge Torres Gómez, Jennifer Simonjan and Falko Dressler, "Low-Complex Synchronization Method for Intra-Body Links in the Terahertz Band," IEEE Journal on Selected Areas in Communications, vol. 42 (8), pp. 1967–1977, August 2024.
        \item \textbf{Paper D:} Jorge Torres Gómez, Jennifer Simonjan, Josep Miquel Jornet and Falko Dressler, "Optimizing Terahertz Communication Between Nanosensors in the Human Cardiovascular System and External Gateways," IEEE Communications Letters, vol. 27 (9), pp. 2318–2322, September 2023.
        \item \textbf{Paper E:} Jorge Torres Gómez, Joana Angjo and Falko Dressler, "Age of Information-based Performance of Ultrasonic Communication Channels for Nanosensor-to-Gateway Communication," IEEE Transactions on Molecular, Biological and Multi-Scale Communications, vol. 9 (2), pp. 112–123, June 2023.
    \end{itemize}
    \item In this chapter, we formulate closed-form expressions for the random variables $T_g$, the average of $T_d$, and the average \ac{PAoI} metric.
    We also display and discuss the resulting metric.
    This chapter is related to the contributions:
    \begin{itemize}
        \item \textbf{Paper E:} Jorge Torres Gómez, Joana Angjo and Falko Dressler, "Age of Information-based Performance of Ultrasonic Communication Channels for Nanosensor-to-Gateway Communication," IEEE Transactions on Molecular, Biological and Multi-Scale Communications, vol. 9 (2), pp. 112–123, June 2023.
        \item \textbf{Paper F:} Jorge Torres Gómez, Ketki Pitke, Lukas Stratmann and Falko Dressler, "Age of Information in Molecular Communication Channels," Elsevier Digital Signal Processing, Special Issue on Molecular Communication, vol. 124, pp. 103108, May 2022.
    \end{itemize}
\end{enumerate}
Finally, we include a \textbf{chapter~5} summarizing future research directions in this topic.

%

\section{Conclusions}

The \ac{AoI} concept conveys the theoretical framework to describe the information freshness in \ac{IoBNT} networks.
Specifically, the average \ac{PAoI} metric is utilized to assess the monitor update period to display new information status.
With the suited interpretation of the information flow within the vessel networks, this metric is calculated with the concentration of nanosensors in the infection location, their path from the infection to the gateway, and their communication capabilities with the gateway.
These random events are all enclosed within the \ac{AoI} concept, leading to a single closed-form expression for the average \ac{PAoI} metric.
As a result, this metric comprehensively evaluates the monitoring capabilities of the \ac{IoBNT} network and ultimately addresses the question of time sensitivity.

%

\chapter[Nanosensors Mobility Model]{Nanosensors Mobility Model along the Cardiovascular System}
\label{sec_mobility}

\section{Introduction}

This chapter introduces nanosensor mobility in the human cardiovascular system.
As the nanosensors are the information carriers within the \ac{IoBNT} framework (described in \Cref{sec_introduction}), the mobility model will ultimately define the transmission delay ($T_d$) and also the generation time ($T_g$).
We model the nanosensor mobility on the macroscopic scale.
This scale refers to locating the nanosensor with the vessel segment rather than the exact position within the vessel segment (microscopic scale).
This is akin to a rough estimate of the mobility in 1D versus 3D model coordinates, pursuing a simple model for the already complex network of the cardiovascular system.\footnote{More accurate models are part of future work, where the random mobility within the human vessels is also included to evaluate the transition among Markov states.}

The model follows from the reported studies in \cite{brown1993bayesian}, where the various part of the cardiovascular system is regarded as a connection of modules along closed circuits; see \cite[Fig. 1]{brown1993bayesian}.
We extended this model with a Markov model formulation where each vessel segment is abstracted as a state of the Markov model, and the transition from one segment to another with the transition probability among states.
That is, we look at the nanosensor's position as a state (vessel segment), and the Markov model can be used to describe the next position of the nanosensor with a given probability.

We remark on the adequacy of the nanosensor's mobility and the Markovian assumption (see \cite[Eq. (1.1.3)]{howard1971dynamic}) with the mechanics of the nanosensor's mobility along the vessels.
The probability that the nanosensor will move to its next position depends only on its current state, not its previous state.
For instance, once a nanosensor is located on the arcus aorta, next to the left ventricle of the heart, it jumps to the next vessel segment (Carotis Communis, Subclavia, or Thoratica) irrespective of the previous nanosensor position (coming from the legs, center body, arms, or the head).

%

\section{Markov Model of the Human Cardiovascular System}

\begin{figure}
  \includegraphics[width=\linewidth]{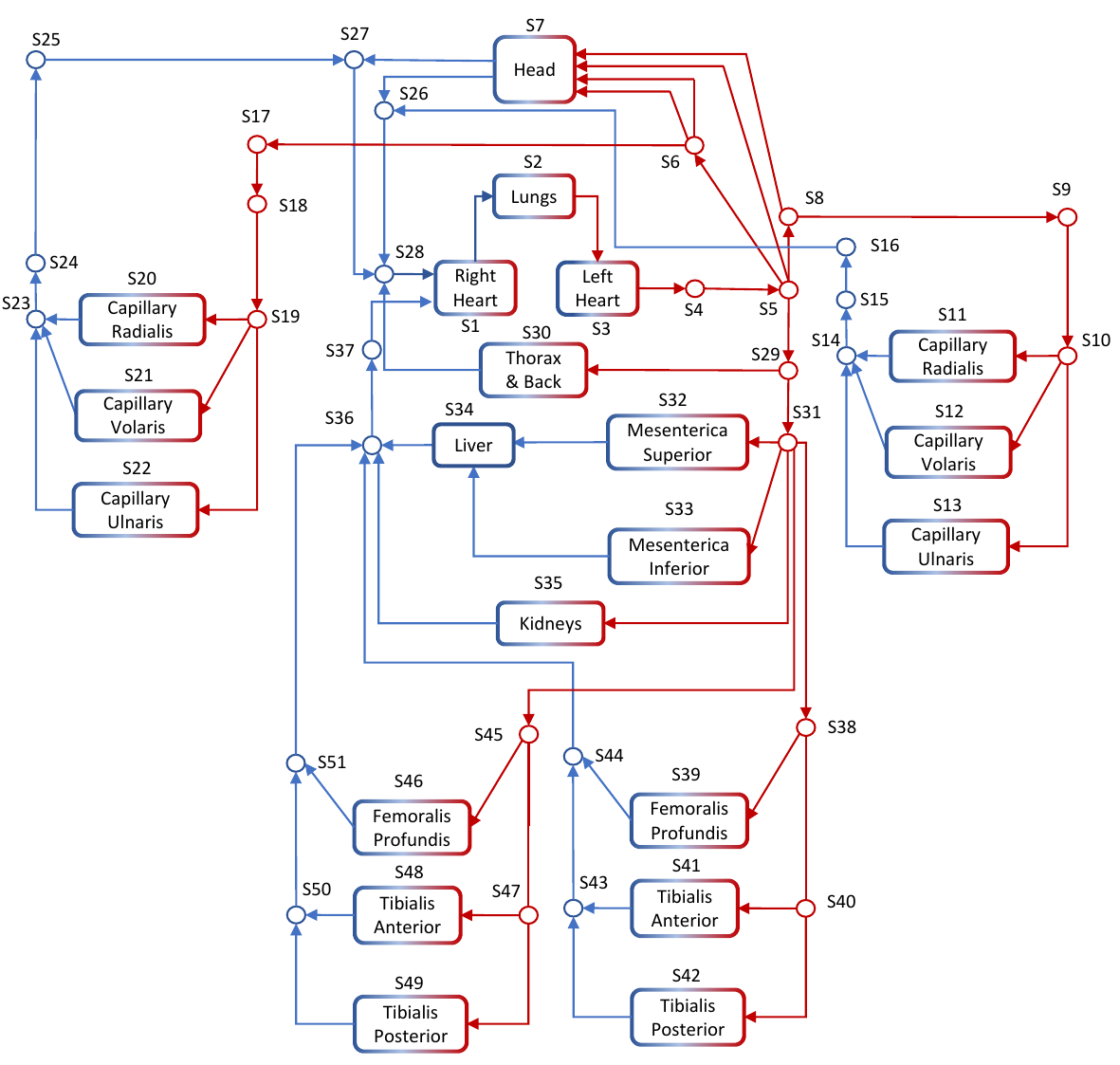}
  \caption{Markov model of the human cardiovascular system \cite{torres-gomez2024electric}.}
  \label{fig_Markov}
\end{figure}

Our Markov model mirrors the most relevant arteries, capillaries, and veins in the human cardiovascular system, as represented in \Cref{fig_Markov}.
States, denoted in the figure with S-labels, directly map the following
\begin{itemize}
    \item Capillaries: These are the thinnest vessel segments; their main function is to bring nutrients to the main tissues and collect their toxic components resulting from metabolic processes of cells.
    In this model, capillaries are in total \num{15} as listed within the last column in \Cref{tab_capillary}.
    The table lists the left part of the body and similarly follows for the right part.
    The capillaries are represented in the Markov model with boxes with the corresponding label as indicated within the second column in the table; see an example with the annotations in \Cref{fig_Markov} for the Right Heart (S1), Lungs (S2), and Left Heart (S3).
    \item Arteries: These vessels transport nutrients and oxygen to the capillaries.
    These can be found in the second column of \Cref{tab_arteries}, totaling \num{25} for the left side of the body.
    A similar annotation follows for the right part of the body.
    The Markov model depicts the arteries with connections between the states, which are shown as red circles. 
    For instance, the connection S4 to S5, next to the left heart block in \Cref{fig_Markov} represents the Arcus aorta, as indicated in the entry \num{12} in \Cref{tab_arteries}.
    \item Veins: These are the vessel segment that connects the capillaries to the Right Heart compartment.
    They are not listed in a table here (the literature mainly focuses on the arteries), but are illustrated within the connections among the circles in blue. 
    For instance, the connection between circles S26 and S38 (left to the Right Heart block in \Cref{fig_Markov}) corresponds to the Superior Vena Cava.
\end{itemize}

The states in the Markov model directly reflect the nanosensor's current position, i.e., vessel segments.
The transition among states directly reflects the nanosensors mobility pattern.
For instance, a node located at the {Arcus} aorta would reflect its state as S5, and from there, the nanosensor might transition with some probability to the head (S7) or the left arm through the {Subclavia} aorta (S9).
In this way, the Markov model representation in \Cref{fig_Markov} mirrors the possible pathways for a nanosensor in the human cardiovascular system.

\begin{center}
    \begin{table}[htbp]
  \caption{Markov states and capillaries.}
    \begin{tabular}{rccl}
          & \multicolumn{1}{l}{Body Part} & \multicolumn{1}{l}{Markov State} & Capillary \bigstrut[b]\\
    \hline
    1     &       & S7  & Head \bigstrut\\
    \hline
    2     & \multicolumn{1}{c}{\multirow{3}[6]{*}{\begin{sideways}Arms\end{sideways}}} & S11 & Radialis \bigstrut\\
    \cline{3-4}    3     &       & S12 & Volaris \bigstrut\\
    \cline{3-4}    4     &       & S13 & Ulnaris \bigstrut\\
        \hline
        5     & \multirow{8}[16]{*}{\begin{sideways}Center Body\end{sideways}} & S1  & Right Heart \bigstrut\\
    \cline{3-4}    6     &       & S2  & Lungs \bigstrut\\
    \cline{3-4}    7     &       & S3  & Left Heart \bigstrut\\
    \cline{3-4}    8     &       & S30 & Thorax and Back \bigstrut\\
    \cline{3-4}    9     &       & S32 & Mesenterica Superior \bigstrut\\
    \cline{3-4}    10    &       & S33 & Mesenterica Inferior \bigstrut\\
    \cline{3-4}    11    &       & S34 & Liver \bigstrut\\
    \cline{3-4}    12    &       & S35 & Kidneys \bigstrut\\
        \hline
        13    & \multicolumn{1}{c}{\multirow{3}[6]{*}{\begin{sideways}Legs\end{sideways}}} & S39 & Femoralis Profundis \bigstrut\\
    \cline{3-4}    14    &       & S41 & Tibialis Anterior \bigstrut\\
    \cline{3-4}    15    &       & S42 & Tibialis Posterior \bigstrut\\
    \hline
    \end{tabular}%
    \begin{tablenotes}
        \item As for the arms and the legs, we list in this table the left side only.
        The right side follows the same names as for the capillary systems.
    \end{tablenotes}
  \label{tab_capillary}%
\end{table}%
\end{center}

\begin{table}[htbp]
  \centering
  \caption{Markov states and arteries.}
    \begin{tabular}{rccr}
          & \multicolumn{1}{l}{Body Part} & \multicolumn{1}{l}{Markov State} & \multicolumn{1}{l}{Capillary} \bigstrut[b]\\
    \hline
    1     & \multirow{4}[8]{*}{\begin{sideways}Head\end{sideways}} & \multirow{2}[4]{*}{S6 to S7} & \multicolumn{1}{l}{Carotis Communis dixtra} \bigstrut\\
\cline{4-4}    2     &       &       & \multicolumn{1}{l}{Vertebralis dixtra.} \bigstrut\\
\cline{3-4}    3     &       & \multirow{2}[4]{*}{S8 to S7} & \multicolumn{1}{l}{Carotis Communis} \bigstrut\\
\cline{4-4}    4     &       &       & \multicolumn{1}{l}{Vertebralis} \bigstrut\\
    \hline
    5     & \multicolumn{1}{c}{\multirow{6}[12]{*}{\begin{sideways}Arms\end{sideways}}} & S6 to S17 & \multicolumn{1}{l}{Subclavia sinistra} \bigstrut\\
\cline{3-4}    6     &       & S8 to S9 & \multicolumn{1}{l}{Subclavia dixtra} \bigstrut\\
\cline{3-4}    7     &       & S9 to S10 & \multicolumn{1}{l}{Axilaris} \bigstrut\\
\cline{3-4}    8     &       & S9 to S11 & \multicolumn{1}{l}{Brachialis} \bigstrut\\
\cline{3-4}    9     &       & S10 to S12 & \multicolumn{1}{l}{Radialis} \bigstrut\\
\cline{3-4}    10    &       & S10 to S13 & \multicolumn{1}{l}{Ulnaris} \bigstrut\\
    \hline
    11    & \multirow{11}[22]{*}{\begin{sideways}Center Body\end{sideways}} & S3 to S4 & \multicolumn{1}{l}{Ascendens} \bigstrut\\
\cline{3-4}    12    &       & S4 to S5 & \multicolumn{1}{l}{\multirow{2}[4]{*}{Arcus}} \bigstrut\\
\cline{3-3}    13    &       & S5 to S8 &  \bigstrut\\
\cline{3-4}    14    &       & S5 to S29 & \multicolumn{1}{l}{\multirow{3}[6]{*}{Thoratica}} \bigstrut\\
\cline{3-3}    15    &       & S29 to S31 &  \bigstrut\\
\cline{3-3}    16    &       & S31 to S38 &  \bigstrut\\
\cline{3-4}    17    &       & S29 to S30 & \multicolumn{1}{l}{Coelica} \bigstrut\\
\cline{3-4}    18    &       & S29 to S31 & \multicolumn{1}{l}{Adominalis} \bigstrut\\
\cline{3-4}    19    &       & S31 to S32 & \multicolumn{1}{l}{Mesenterica Superior} \bigstrut\\
\cline{3-4}    20    &       & S31 to S33 & \multicolumn{1}{l}{Mesenterica Inferior} \bigstrut\\
\cline{3-4}    21    &       & S31 to S35 & \multicolumn{1}{l}{Renalis} \bigstrut\\
    \hline
    22    & \multicolumn{1}{c}{\multirow{4}[8]{*}{\begin{sideways}Legs\end{sideways}}} & S38 to S39 & \multicolumn{1}{l}{Femoralis} \bigstrut\\
\cline{3-4}    23    &       & S38 to S40 & \multicolumn{1}{l}{Poplitea} \bigstrut\\
\cline{3-4}    24    &       & S40 to S41 & \multicolumn{1}{l}{\multirow{2}[4]{*}{Tibialis}} \bigstrut\\
\cline{3-3}    25    &       & S41 to S42 &  \bigstrut\\
    \hline
    \end{tabular}%
  \label{tab_arteries}%
\end{table}%

A primary advantage of the Markov model is its mathematical tractability that follows from linear algebra.
The model is fully described with the transition probability matrix, as
\begin{equation}\label{eq_P_PI}
    \Pi=\left[ \begin{matrix}
        p_\text{S1 S1} & p_\text{S1 S2} & \cdots & p_\text{S1 SN}\\
        p_\text{S2 S1} & p_\text{S2 S2} & \cdots & p_\text{S2 SN}\\
        \vdots &  & & \\
        p_\text{SN S1} & p_\text{SN S2} & \cdots & p_\text{SN SN}\\
    \end{matrix}
    \right],
\end{equation}
where $\text{N}=51$ is the number of states and $p_\text{Si Sj}$ refers to the transition probabilities between the states pair \text{Si} to \text{Sj}; we discuss means to evaluate these transition probabilities in the next section.

Having the \Cref{eq_P_PI}, we can evaluate the time evolution for the nanosensor position.
That is, we can evaluate probabilities such as the nanosensor to follow a path trajectory from the infection location to the gateway and the likelihood of finding a nanosensor within the infection location.
These are two probabilities needed later to evaluate the average \ac{PAoI} metric. Precisely, these probabilities are the expected value of the generation time ($T_g$) and the delay ($T_d$), as described in steps \num{3} and \num{4} of our strategy in \Cref{sec_strategy}.

Related to the probability of finding a nanosensor at the infection location within an arbitrary vessel segment, we evaluate the stationary probability vector $\boldsymbol{\nu}$ solving for the equation 
\begin{equation}\label{eq_stationary}
    \boldsymbol{\nu}=\Pi\boldsymbol{\nu},
\end{equation}
which follows from see \cite[Eq. (1.3.16)]{howard1971dynamic}, where the components of $\boldsymbol{\nu}$, as ${\nu_i}$, yield the probability of finding a single nanosensor within the vessel segment $i$.

Calculating the nanosensor's probability of traveling from the infection location to the gateway is less expeditious but straightforward.
Once the nanosensor is in the infection location, it travels with probability \num{1} to the heart, as all the veins sink on the right ventricle; see \cite[Fig. 1-1]{guyton2015guyton}.
Once in the heart, the nanosensor might travel directly to the gateway location, here denoted with probability $p_{s=\mathrm{G}}$, or it might wander to another trajectory first, evaluated as $p_{s=\mathrm{G}}\times(1-p_{s=\mathrm{I}}-p_{s=\mathrm{G}})$, where $p_{s=\mathrm{I}}$ denotes the probability for the nanosensor to travel from the heart to the infection location again.
In this calculation, we exclude the probability of the nanosensor traveling twice to the infection location, as this would account for a different sample.
Then, if we consider that the nanosensor might travel an infinite amount of times to the capillaries other than the infection and the gateway, the probability for the nanosensor to travel from the infection to the gateway yields
\begin{align}\label{eq_transition}
   p_\mathrm{I \to G}=&p_{s=\mathrm{G}}\times \sum_{k=0}^\infty{(1-p_{s=\mathrm{I}}-p_{s=\mathrm{G}})^k},\\
   &=\frac{p_{s=\mathrm{G}}}{p_{s=\mathrm{I}}+p_{s=\mathrm{G}}},\nonumber
\end{align}
after using the MacLaurin series formula $\sum_{k=0}^\infty{(1-p)}^k=\frac{1}{p}$.

The probabilities $p_{s=\mathrm{G}}$ and $p_{s=\mathrm{I}}$ can be directly evaluated with the transition probabilities from the left ventricle in the heart (state $s_3$ in \Cref{fig_Markov}) to the gateway or the infection location, respectively.
For instance, assuming the infection is on the shoulder (Clavialis), the probability $p_{s=\mathrm{I}}$ readily yields the product $(p_{s3,s4}\times p_{s4,s5}\times p_{s5,s8})$, and similar evaluation is performed for other locations.
The following section details the evaluation for the transition probabilities in \Cref{eq_P_PI}.

%

\section[Evaluation of the Transition Probabilities]{Evaluation of the Transition Probabilities}
\label{sec_frankesntein}

\begin{wrapfigure}{R}{0.4\textwidth}
    \centering
    \includegraphics[width=\linewidth]{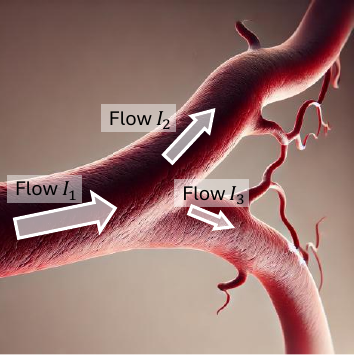}
    \caption{Representation of flows within the vessels bifurcation.}
    \label{fig_bifurcation}
\end{wrapfigure}

The calculations above in Eqs. \eqref{eq_stationary} and \eqref{eq_transition} ultimately depend on the transition probabilities among states, which follow the physiology of the human vessels.
Once the nanosensor faces a bifurcation in the human cardiovascular system, it follows one direction or the other with the blood flow.
We hypothesize that this probability relies only on the flow ratio from the current vessel segment to the following segments; see \Cref{fig_bifurcation}.
Following the representation in \Cref{fig_bifurcation}, the transition probabilities can be evaluated as \footnote{This hypothesis is based on the radial symmetry of particles observed within the blood flow; see \cite{lee2013near}.}

\begin{equation}\label{eq_transition_2}
    p_{s_{1},s_{2}}=\frac{I_{2}}{I_{1}},
\end{equation}

Following this relation, the transition probabilities are evaluated by the flow on each vessel segment.
We develop a simulator of the human arteries to numerically evaluate the variety of flows and, as such, the transition probabilities.
The following subsection provides details on the simulator.

\begin{figure}[t]
\centering
\includegraphics[width=\linewidth]{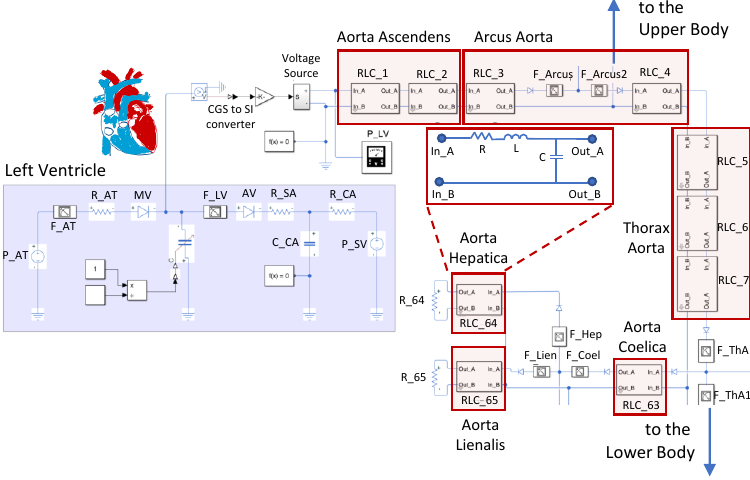}
\caption{Electric circuit design of the human arteries following the design in \cite{noordergraaf1963use}.}
\label{fig_simulink}
\end{figure}   

\subsection{Electric Circuit Simulator of the Human Cardiovascular System}

We introduced with the work in \cite{torres-gomez2024electric} a simulator for the cardiovascular system where vessel segments are modeled through an electric circuit, as first suggested by \textcite{westerhof1969analog}.
\Cref{fig_simulink} illustrates a portion of the model, including a block for the left ventricle and various blocks representing vessel segments in the arteries. 
This figure explicitly illustrates blocks for the aortas Ascendes, Arcus, Thorax, Coelica, Lienalis, and Hepatica.
In this model, all the vessel segments connect a resistor, a capacitor, and an inductor following an L-inverted topology, as illustrated for the Aorta Hepatica in \Cref{fig_simulink}. 

The resistor models the vessels' resistance to the flow, the capacitor models the elasticity of the human vessels, and the inductor models the inertia of the bloodstream.
The voltage in the circuit is directly related to blood pressure and the current of blood flow.\footnote{The design for the electric circuit representation is accessible in \url{https://github.com/tkn-tub/frankenstein}}
These circuit components are evaluated for each vessel segment, according to its length $\Delta l$, radius $R_v$,  and thickness $h$, as \cite{rideout1991mathematical}
\begin{equation}\label{eq:RLC}
    R=\frac{8\pi\mu \Delta l}{\pi R_v^2},\ L=\frac{9\rho\Delta l}{4\pi R_v^2},\ C=\frac{3\pi R_v^3\Delta l}{2Eh},
\end{equation}
where $\mu$ is the blood viscosity, $\rho$ is the blood density, and $E$ is the Young's bulk modulus of elasticity.
We calculate the resistor, inductor, and capacitor values according to the physiological values reported in \cite{westerhof1969analog} for the main arteries.\footnote{The complete calculation is given in the repository \url{https://github.com/tkn-tub/frankenstein}}

\begin{figure}[!h]
\centering
\includegraphics[width=\linewidth]{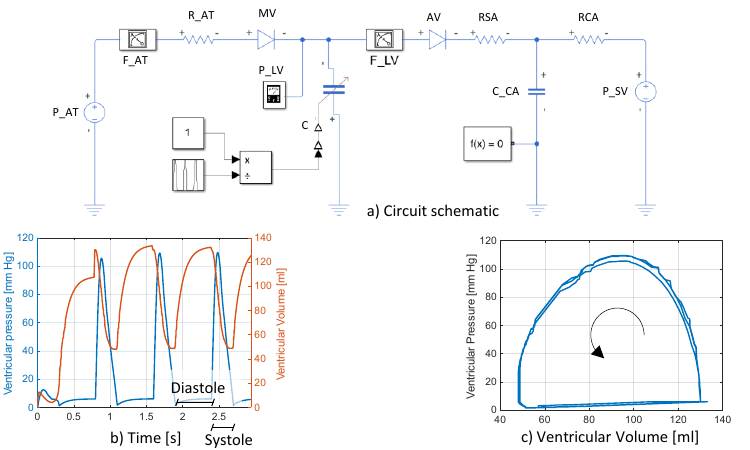} 
\caption{Electric circuit for the heart \cite{rideout1991mathematical} and resulting pressures and cardiac cycle in the left ventricle \cite{torres-gomez2024electric}.}
\label{fig_circuit_heart}
\end{figure}

As for the heart, we introduce the schematic model in \Cref{fig_circuit_heart}~a), which comprises the connection of two main parts: 
\begin{itemize}
    \item The left ventricle, which includes the voltage source $\mathrm{P\_}\mathrm{AT}$, the resistor $\mathrm{R\_}\mathrm{AT}$, the diodes $\mathrm{MV}$ and $\mathrm{AV}$, as well as the variable capacitor $\mathrm{C}$.    
    \item The systemic circulation, which includes the T-topology of the resistors $\mathrm{RSA}$ and $\mathrm{RCA}$, with the capacitor $\mathrm{C\_}\mathrm{CA}$.
    The topology is terminated with the source voltage $\mathrm{P}_\mathrm{SV}$, which accounts for central venous pressure.
\end{itemize}

The heartbeat is modeled with the variable capacitor $\mathrm{C}$, recreating the diastole and systole period of the heart; see \Cref{fig_circuit_heart}~b).
The capacitor is modulated with an oscillating signal (clipped sine wave) of frequency $f_\mathrm{heart}$; see the connection in \Cref{fig_circuit_heart}~a).\footnote{Details on this signal are given in \cite[Sec. 3.1]{torres-gomez2024electric}.}
The voltage source $\mathrm{P\_}\mathrm{AT}$ provides the atrial pressure, and diodes set the proper direction of currents with the physiology of the heart.\footnote{Due to the resemblance of this simulator with the development of the main character in the novel Frankenstein, we follow the same name for this project.
We provide open access to the code in the repository \url{https://github.com/tkn-tub/frankenstein}}
This model for the heart follows the design in \cite[Fig. 4.2.1 page 79]{rideout1991mathematical} and exhibits a good correspondence of ventricular pressure (\Cref{fig_circuit_heart}~b)) and cardiac cycle (\Cref{fig_circuit_heart}~c)) with the reported literature in hemodynamic; see \cite[Fig. 9.8 page 118]{guyton2015guyton}.

\begin{figure}[t]
\centering
\includegraphics[width=0.8\columnwidth]{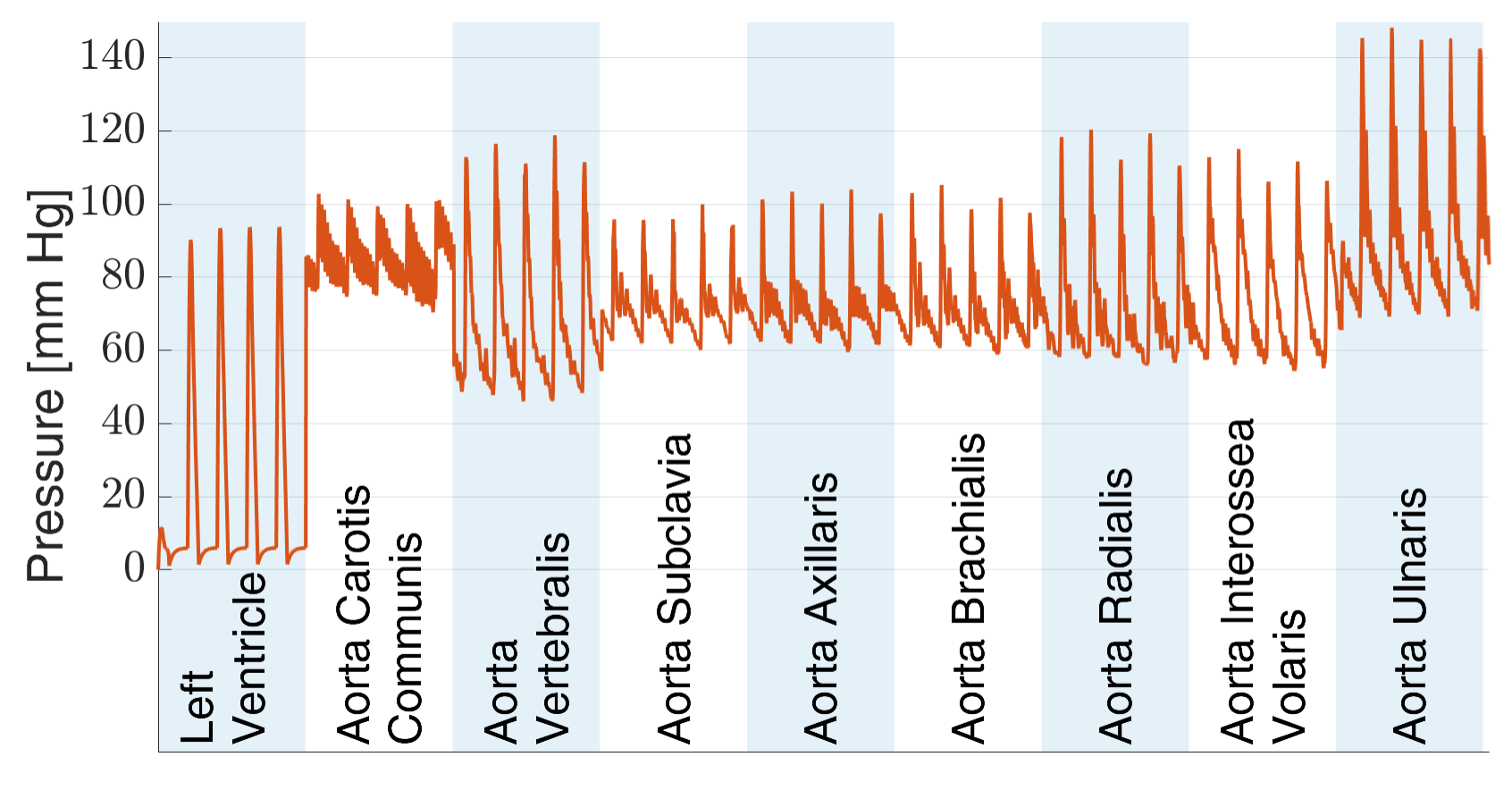} 
\caption{Blood pressure in head and arms for the simulator in \cite{torres-gomez2024electric}.}
\label{fig_blood_pressure_Arms}
\end{figure}

\subsection{Results}

The simulator demonstrates results for pressure and flow that closely align with studies in physiology books; the pressure for the upper arms in \Cref{fig_blood_pressure_Arms} resembles the pressure pattern printed in \cite[Fig. 14-2 page 172]{guyton2015guyton}.
With this simulator, we can finally evaluate the flows at the bifurcations and, accordingly, the transition probabilities among states in \Cref{eq_transition_2}.
Using the evaluated transition probabilities, we compute the transition matrix in \Cref{eq_P_PI} and numerically solve for the stationary probabilities in \Cref{eq_stationary}. 

\begin{figure}
\centering
\includegraphics[width=\linewidth]{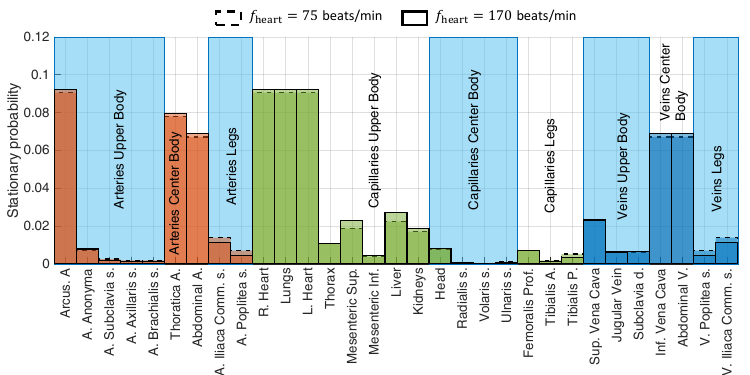} 
\caption{Distribution of nanosensors in the cardiovascular system when the body is at rest $f_\mathrm{heart}=\SI{75}{\beat\per\min}$ and when doing sports \mbox{$f_\mathrm{heart}=\SI{170}{\beat\per\min}$} \cite{torres-gomez2024electric}.}
\label{fig_results_nanosensors_distribution}
\end{figure}

\Cref{fig_results_nanosensors_distribution} illustrates the results of evaluating the stationary probabilities for various vessel segments.
The largest probability is in the center organs (Heart and Lungs) and the arteries and veins with the most significant dimensions (Arcus, Thoracic, and Vena Cava).
The plot also illustrates the variability of this probability with the activity, which can be modeled with the frequency of the heart.
With the activity, this probability will slightly increase in some segments and decrease in others; it mostly increases in the capillaries within the center body, i.e., the Mesenteric and Liver.

%

\section{Conclusions}

The Markov model stands pragmatic as it submits a single-parameter framework to examine mobility along the cardiovascular system's network.
The parameter is given with the transition probability matrix, enabling the evaluation of the essential metrics: the concentration of nanosensors with the stationary vector and the likelihood of following a pre-defined path along the vessels.
The simplicity of the Markov model is also followed by the simplicity of the electric circuit simulator in evaluating the transition probabilities matrix.
The simplicity of the simulator stems from the vessel's division into segments, which are all modeled with the same topology as the RLC circuit.
Incorporating this circuit into the Markov model effectively captures the large-scale mobility of nanosensors with low-complexity calculations.
Furthermore, it allows mimicking the hemodynamics of specific subjects as a digital twin model; we explore this direction further in the \Cref{sec_future}.

%

\chapter{Modeling the Nanosensor-to-Gateway Communication Link}
\label{sec_comm}

\section{Introduction}

This chapter summarizes two widespread models for the communication link between nanosensors and the gateway.
We describe the ultrasonic channel, which has the most extensive communication range, and the terahertz channel, which has the lower range but a smaller size for the transducers.
See examples of technological development as ultrasonic-based implantable devices in \cite{santagati2020design} and the potential of functionalized nanoantennas in \cite{sangwan2022joint}.

\begin{wrapfigure}{R}{0.6\linewidth}
	\centering
	\includegraphics[width=\linewidth]{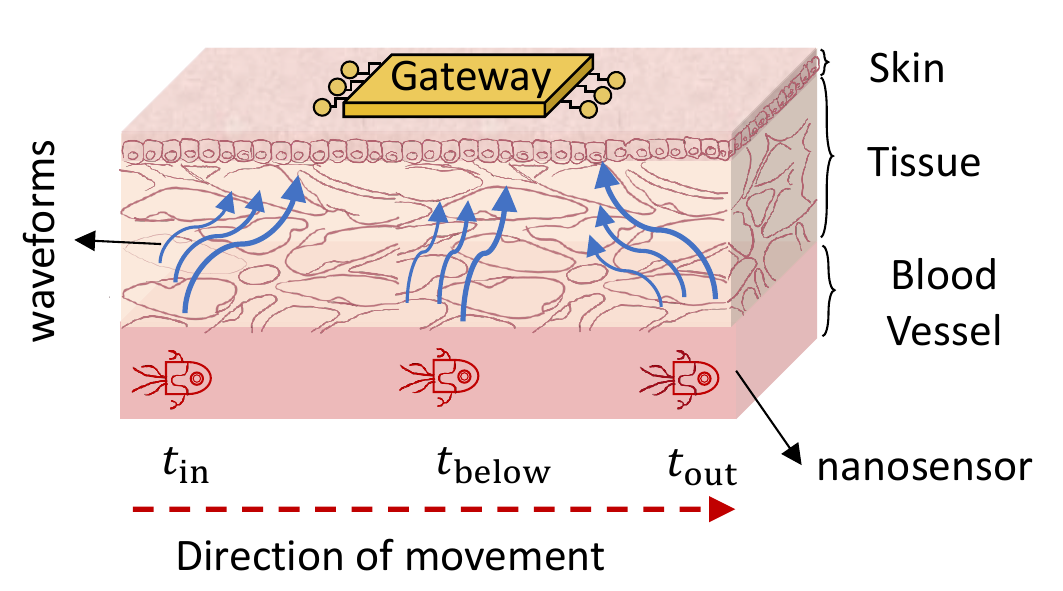}
	\caption{Conceptual representation of the nanosensor and gateway connection.}%
	\label{fig_system_model}
\end{wrapfigure}

The evaluation of the communication link is abstracted with the rate of failed deliveries, related to the Random event~3 in \Cref{sec_strategy}, and feeding the term $p_\mathrm{loos}$ in the average \ac{PAoI} concept.
The corresponding metrics are the \ac{BER} for binary updates and \ac{PER} when the updates consist of multiple bits to represent the sample.

Both channel models are evaluated with the scenario illustrated in \Cref{fig_system_model}, where the nanosensor flows straight from left to right in the blood vessel.
This assumption is valid in vessel segments where the flow is laminar and when advection dominates diffusion \cite{jamali2019channel}. 
The variables $t_i$ mark specific points in time with the mobility of the nanosensor: Timepoint $t_\text{in}$ indicates when the nanosensor gets into the communication range of the gateway, $t_\text{below}$ marks when the nanosensor is directly below the gateway, and $t_\text{out}$ when the nanosensor leaves the communication zone with the gateway.

To evaluate the communication performance, we describe mainly two components of the received signal at the gateway: i) the power level and ii) the distortion.
The power level deals with the channel attenuation when the signal propagates through the body tissues; see a representation in \Cref{fig_system_model}.
The distortion deals with the impact of the nanosensor mobility while driven by the bloodstream.

Due to the low-propagation speed of the ultrasonic waveform, in the magnitude order of meter per millisecond \cite{bos2019enabling}, and the low transmission frequency, in the magnitude order of the megahertz, packets are transmitted along the millisecond timescale.
As a result, packets are highly affected by the nanosensors mobility.
In the case of terahertz transmissions, the high bandwidth affords performing emissions within the channel coherence time, thereby avoiding the impact of nanosensor mobility.
The nanosensor mobility defines a channel coherence time in the millisecond timescale, which sets a transmission bandwidth in the gigahertz range.
In the following two sections, we provide further details on these calculations.

%

\section{Ultrasonic Channel}

The ultrasonic channels refer to a communication scheme where information is carried by ultrasonic waveforms propagating through the tissue.
The channel is modeled as a \ac{LTV} system, see \cite{torres-gomez2022modeling}, where the received signal is given as \cite{matz2011fundamentals} 
\begin{equation}\label{eq_rx_signal}
    r(t)=g_ds(t-\tau_d)e^{j2\pi\nu t},
\end{equation}
and $s(t)$, $g_d$, $\tau_d$, and $\nu$ are the transmitted signal, gain factor, delay, and Doppler frequency, respectively.
The subscript~$d$ in the above expression denotes the distance dependency of the gain and the delay.

\begin{wrapfigure}{R}{0.5\linewidth}
	\centering
	\includegraphics[width=\linewidth]{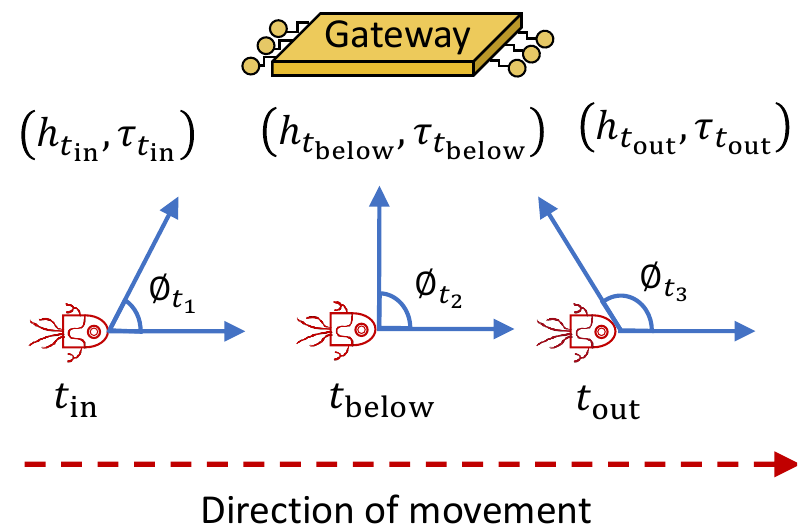}
	\caption{Illustration of parameters for the ultrasonic channel model \cite{torres-gomez2022modeling}.}
	\label{fig_system_model_2}
\end{wrapfigure}

The Doppler frequency term is given as \cite{matz2011fundamentals}
\begin{equation}\label{eq_Doppler}
    \nu=v\cos(\phi)\frac{f_c}{c_u},
\end{equation}
where $v$ is the velocity of the nanosensor, $c_u=\SI{1480}{\meter\per\second}$ is the propagation speed of the ultrasonic waveform, $f_c=\SI{1}{\mega\hertz}$ is the carrier frequency, and $\phi$ is the angle of arrival of the wave relative to the direction of motion of the nanosensor, as depicted in \Cref{fig_system_model_2}.
These speed and center frequency values follow the reported ones in~\cite{bos2019enabling}.
For the in-body scenario, we consider typical velocities $v$ of the nanosensor in the human cardiovascular system in the aorta ($v=\SI{0.2}{\meter\per\second}$), arteries ($v=\SI{0.1}{\meter\per\second}$), and veins ($v=\SI{0.03}{\meter\per\second}$)~\cite{guyton2015guyton}.

At this point, we realize that the model of the received signal in Eqs. \eqref{eq_rx_signal} and \eqref{eq_Doppler} is fully described with the parameters $g_d$, $\tau_d$, and $\nu$, and we evaluate these parameters with the reported measurements in~\cite{bos2019enabling}.
The gain and delay are evaluated with the interpolate results of \ac{CIR} measurements given by~\textcite{bos2019enabling}, where the \ac{CIR} is provided for three specific distances $\SIlist[list-units=single]{20;40;80}{\milli\meter}$ between sender and receiver.\footnote{Published by Thomas Bos under the CC-BY-SA 4.0 license on GitHub \url{https://github.com/BosThomas/USbodyComm}}
These measurements regard the impact of bones and the already included multipath effect.
The gain factor, is readily given with the \ac{RMS} of the given \ac{CIR}, yielding
\begin{equation}\label{eq_att}
    g_{d}=\sqrt{\frac{1}{N}\sum_{k=1}^{N}{h_{d}^2[k]}},
\end{equation}
where $N$ is the length of the sequence and $h_{d}[k]$ is the measured \ac{CIR} at distances $\SIlist[list-units=single]{20;40;80}{\milli\meter}$.

We evaluate the delay ($t_d$) in the ultrasonic communication link as the sum of the mean value of the \ac{CIR} and its standard deviation, both considered w.r.t. the time dimension.
Looking at the \ac{CIR} as a time-dependent distribution of received energy, its mean value represents the average time at which the pulse is received, while its standard deviation indicates the pulse's width \cite{peebles1987probability}.
Following this description, the delay is calculated by adding the mean time to half of the width of the \ac{CIR} peak, yielding
\begin{equation}\label{eq_delay_sonic}
    \tau_{d}=\overline{\tau}_{d}+\frac{\tau_{\sigma,d}}{2},
\end{equation}
where the peak time is by definition
\begin{equation}\label{eq_delay_mean}
    \overline{\tau}_{d}=\frac{1}{f_s}\frac{1}{N}\frac{\sum_{k=1}^{N}{k\cdot h_{d}^2[k]}}{\sum_{k=1}^{N}{h_{d}^2[k]}},
\end{equation}
and the standard deviation is
\begin{equation}\label{eq_delay_std}
    \tau_{\sigma,d}=\frac{1}{f_s}\sqrt{\frac{1}{N}\frac{\sum_{k=1}^{N}{(k-f_s\overline{\tau}_{d})h_{d}^2[k]}}{\sum_{k=1}^{N}{h_{d}^2[k]}}},
\end{equation}
and $f_s$ is the system's sampling rate.

\begin{figure}
	\centering
	\includegraphics[width=0.8\linewidth]{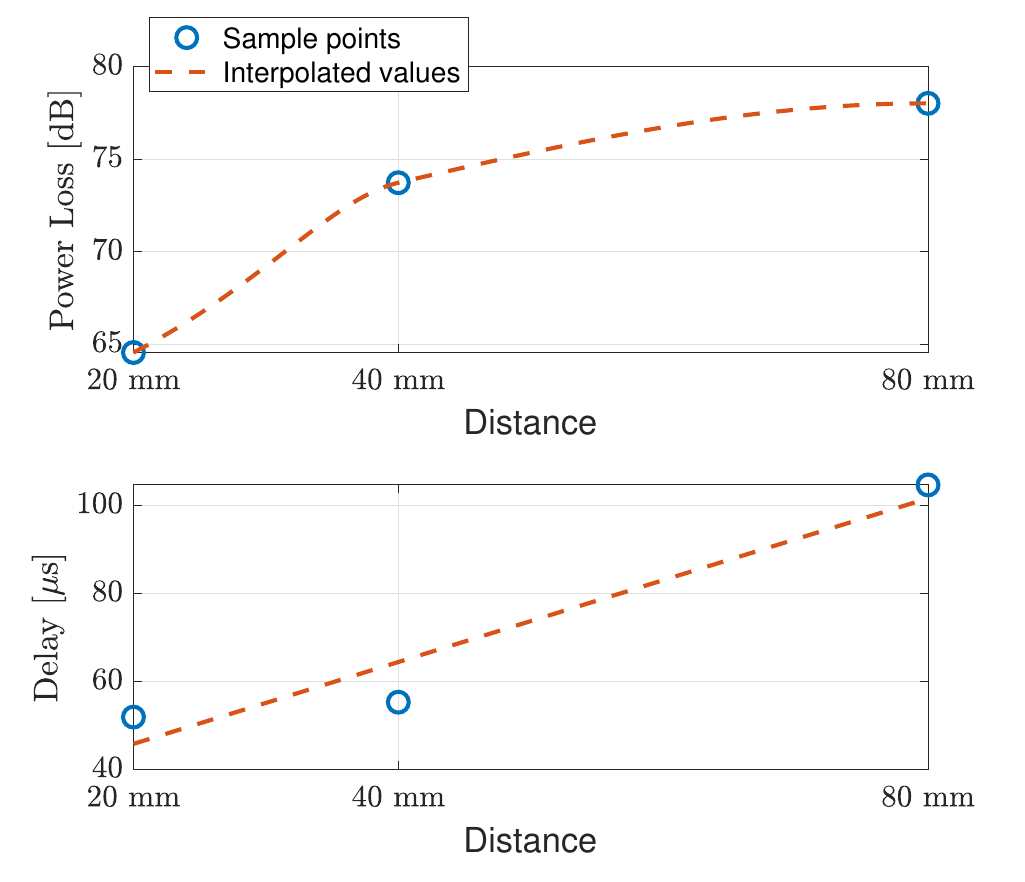}
	\caption{Interpolated power loss and delay over distance for the ultrasonic channel \cite{torres-gomez2022modeling}.}
	\label{fig_res1}
\end{figure}

Next, we interpolate the values provided in Eqs. \eqref{eq_att} and \eqref{eq_delay_sonic} to evaluate the power loss, evaluated units of decibels as $p_L=-20\log_{10}(g_d)$, and the delay ($t_D$) for arbitrary distances.
We use the ``pchip'' cubic interpolation method for the path loss, as it develops a monotonic function for the arguments, i.e., the power loss increases with the distance.
To interpolate the delay, we fit a line to the measured delay because it increases linearly with the distance.
We plot the results of the interpolations in \Cref{fig_res1} with the dashed lines; the blue circles depict the result of evaluating the power loss ($p_L$) and delay ($\tau_d$) with the measurements reported in~\cite{bos2019enabling}.

The Doppler term is readily evaluated with \Cref{eq_Doppler} and following the geometry in \Cref{fig_system_model_2}.
We evaluate these parameters assuming a linear mobility pattern for the nanosensor, depicted in \Cref{fig_system_model_2}, as follows for the laminar flow within the capillaries; see \cite{guyton2015guyton}.
We numerically evaluate the term $\cos(\phi)$ with the trigonometric definition
\begin{equation}
    \cos{\phi}=\frac{d_c}{d},
\end{equation}
where $d_c$ is the distance from the nanosensor to the center (point $t_\mathrm{below}$ in \Cref{fig_system_model_2}) and $d$ is the actual distance from the nanosensor to the gateway.
As for the speed $v$, propagation speed $c_u$, and center frequency $f_c$, we use the values mentioned earlier.

\begin{figure}
	\centering
	\includegraphics[width=0.9\linewidth]{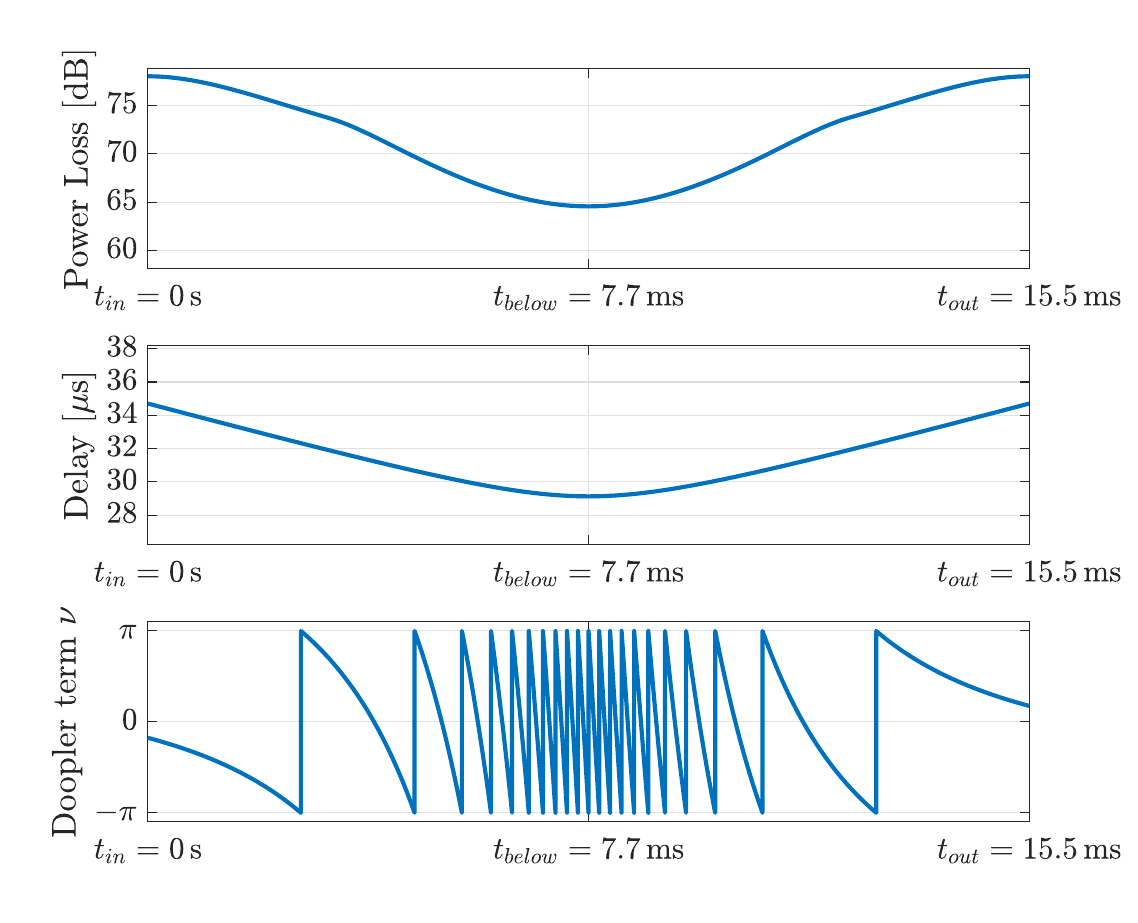}
	\caption{Results on the perceived loss, delay, and the Doppler term on the travel path when the velocity of the nanosensor is $v=\SI{0.2}{\meter\per\sec}$ and the carrier frequency is $f_c=\SI{1}{\mega\hertz}$ \cite{torres-gomez2022modeling}. 
	}
	\label{fig_res2}
\end{figure}

Having the power loss, delay, and Doppler terms for arbitrary distances, we can evaluate those within the path trajectory of the nanosensor; see \Cref{fig_system_model}.
\Cref{fig_res2} depicts these main variables in the nanosensor-to-gateway link when assuming the traveling speed of \SI{0.2}{\meter\per\second} in the aorta.
As the nanosensor displaces straight from left to right, the communication distance to the gateway decreases and increases later, producing a similar pattern for the power loss and the delay; see the results in \Cref{fig_res2}.
The power loss and the delay start at $t_{\mathrm{in}}$ with the highest value, as the nanosensor is at the farthest distance from the gateway.
These variables achieve the minimum at the lowest distance when the nanosensor is just below the gateway at $t_{\mathrm{below}}$, after which it increases again.
However, the Doppler terms $\nu$ introduces a larger time-variability.
This term varies abruptly with the nanosensor position and more markedly when the nanosensor is just below the gateway.
When these variables (power loss, delay, and Doppler) are compared, the Doppler term introduces the highest variability, yielding the most significant distortion on the received sequence at the gateway.

\begin{figure}
	\centering
	\includegraphics[width=0.7\linewidth]{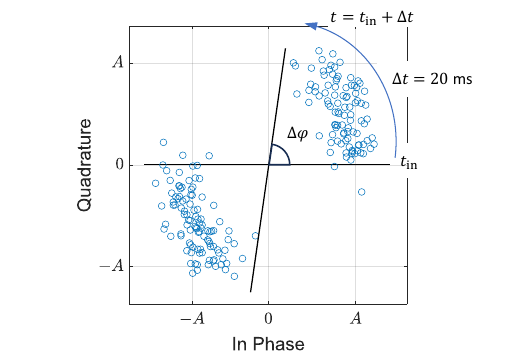}
	\caption{Recovered constellation points.}
	\label{fig_res3}
\end{figure}

\begin{figure}[t]
	\centering
	\includegraphics[width=.8\linewidth]{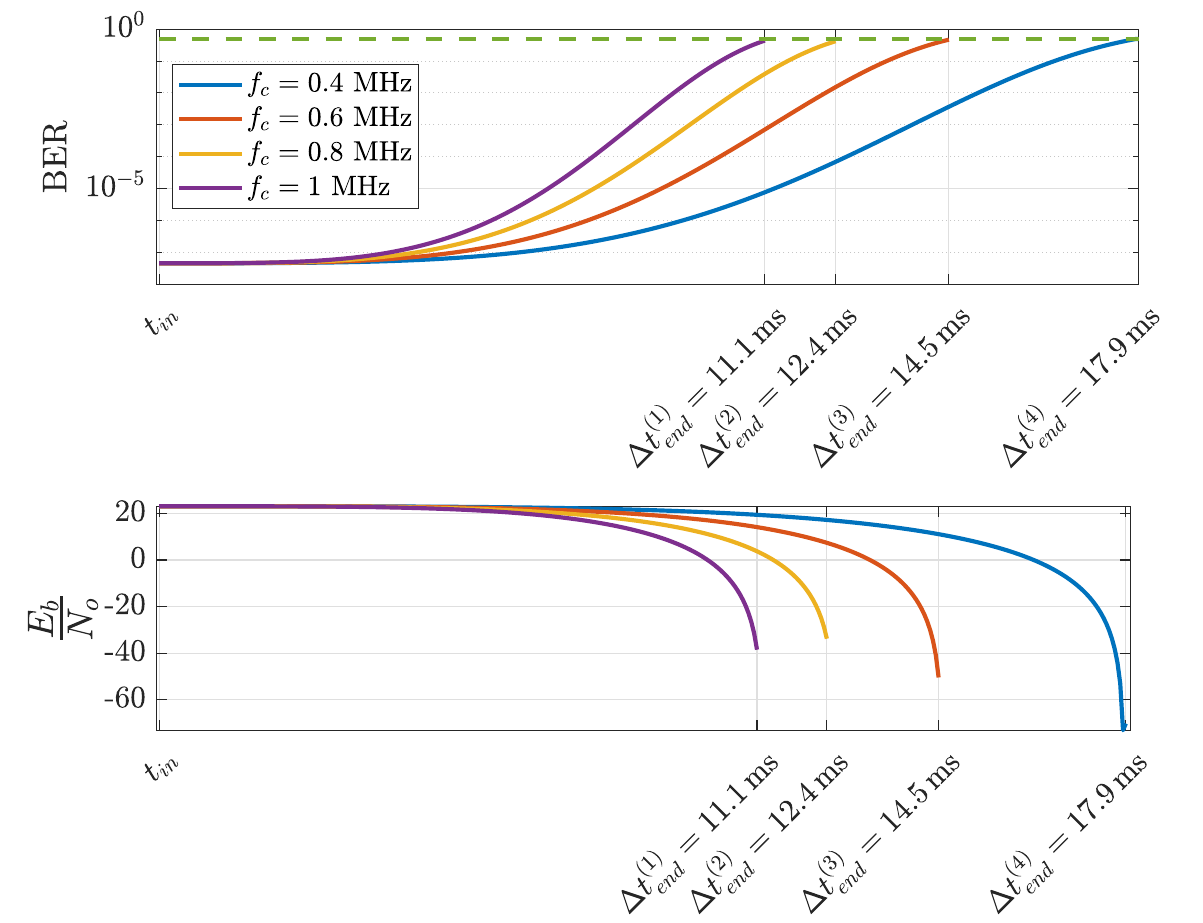}
	\caption{Achievable \ac{BER} versus time when the velocity of the nanosensor is $\SI{0.2}{\meter\per\sec}$ (aorta vessel segment) \cite{torres-gomez2022modeling}.}
\label{fig_res6}
\end{figure}

In more detail, we illustrate in \Cref{fig_res3} the received constellation points from \ac{BPSK} transmissions as the nanosensor displaces from left to right and starting at $t_\mathrm{in}$.
The Doppler term introduces a remarkable distortion on the received waveform as we observe the rotation of the constellation points with time in the amount $\Delta \varphi=2\pi\nu t$.
The impact of the nanosensor mobility severely rotates the original constellation points at $(A,0)$ and $(-A,0)$ counterclockwise, eventually degrading the communication performance on the millisecond time scale.

Finally, to evaluate performance in ultrasonic channels, \Cref{fig_res6} depicts the resulting \ac{BER} and the ratio $\frac{E_b}{N_o}$ with time.
The simulation starts at \mbox{$t_\text{below}=\SI{50}{\milli\second}$}, corresponding to the shortest distance between both at \SI{20}{\milli\meter}, and stop when the \ac{BER} reaches \num{0.5} units.
The plot exhibit the behavior with time till for different center frequencies as $\SIlist{0.4;0.6;0,8;1}{\mega\hertz}$.

From these plots, we can sketch two main conclusions toward a functional design:
\begin{itemize}
    \item The \ac{BER} remains constant with the frequency if the transmission time is less than around $\SI{5}{\milli\second}$.
    That is, the impact of the Doppler is neglible along this time interval.
    \item As a more appealing conclusion, the communication performance degrades in the millisecond timescale.
    That is, the receiver needs to synchronize with the emitter in this timescale.
\end{itemize}

%

\section{Terahertz Channel}

\begin{wrapfigure}{R}{0.6\linewidth}
	\centering
	\includegraphics[width=\linewidth]{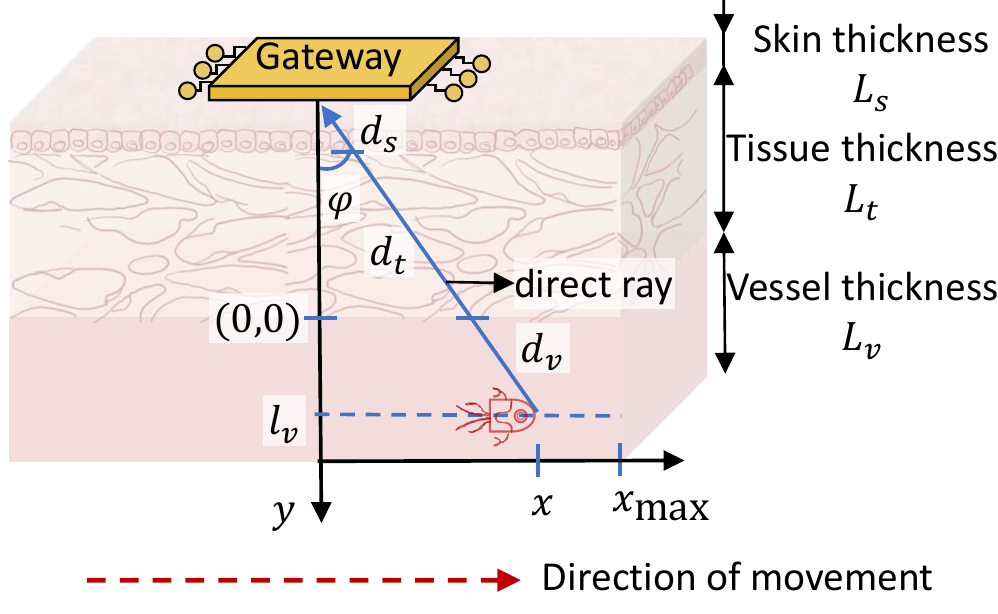}
	\caption{Illustration for the reference coordinate system.}
	\label{fig_system_model_4}
\end{wrapfigure}

Unlike the ultrasonic, the terahertz channel is rather more sensitive to the communication distance.
To frame a point of reference, the power loss in the ultrasonic channel varies in the unit of decibels when the distance varies in the millimeter range.
In the same distance range, the power loss in the terahertz channel varies in the hundreds of decibels; see \cite[Fig. 7]{piro2016terahertz}.
Therefore, it is crucial to detail the nanosensor position within the vessel on a $\si{\milli\meter}$ scale.
In the following sections, we provide a model for the nanosensor dynamic within the vessel and abstract the communication link with the perceived \ac{SNR} parameter, as illustrated in \Cref{fig_system_model_4}.

\subsection{Nanosensor Mobility Model}

In \Cref{fig_system_model_4} we represent the main variables related to the nanosensor mobility and its distance to the gateway.
The coordinates refer to the vertical and horizontal direction only, as represented by the variables $l_v$ and $x$ in the figure.
This representation implicitly accounts for a reduced 2D mobility model of the nanosensor with respect to the Gateway.
This assumption implicitly considers the radiation pattern of the Gateway antenna is constant in the plane vertical to the $xy$-plane in the figure.
A detailed representation of coordinates is given as follows

\begin{itemize}
    \item In the vertical direction:
    \begin{itemize}
        \item The center of reference system, denoted as $(0,0)$, are placed in the vessel wall facing the gateway.
        \item The nanosensor travels in depth within the vessel as denoted with the variable $l_v$, which also indicates the vessel stream.
        \item The thickness of the skin, tissue, and vessel are denoted with the variables $L_s$, $L_t$, and $L_v$, respectively.
    \end{itemize}
    \item In the horizontal direction
    \begin{itemize}
        \item The variable $x_\mathbf{max}$ evaluates the maximum communication range between the nanosensor and the gateway as $2x_\mathrm{max}$.
        \item The variable $x=v_{l_v}t$ denotes the nanosensor position is determined, where $t$ is time, and $v_{l_v}$ is the speed along the stream $l_v$ withing the vessel.
    \end{itemize}  
\end{itemize}
%
The value of $l_v$ depends on the interactions between the nanosensor and the blood fluid before entering the communication segment with the gateway.
For this variable, we assume it follows a uniform distribution in the range $[0,\, L_v]$, here denoted as
\begin{equation}
f_{L_v} \sim \mathcal{U}(0, L_v).    
\end{equation}
Thereby, the nanosensor might be in any particular vessel stream with the same probability.\footnote{This agnostic assumption is founded by the study in \cite{lee2013near}.
However, other distributions might apply here under the size of the nanosensor.}
As the second variable, we evaluate the nanosensor position in the horizontal direction as $x=v_{l_v}t$, see \Cref{fig_system_model}.

%
\subsection{Modeling the Perceived SNR}

As a result of the above mobility model, the experienced \ac{SNR} at the gateway is also a random variable and readily expressed as a function of these two coordinates as
\begin{equation}\label{eq_SNR_1}
    \gamma_\mathrm{m}(l_v)=\frac{P_\mathrm{Tx}}{N_pP_\mathrm{L}(x,l_v)}.
\end{equation}
where $P_\mathrm{Tx}$ is the transmit power of the nanosensor, $N_p$ is the noise power --introduced mainly by the gateway circuit, and $P_\mathrm{L}(x,l_v)$ is the power loss, which exhibits the dependency of the \ac{SNR} with the mobility model ($x$ and $l_v$).
The power loss is evaluated as \cite[Eq. (2)]{simonjan2021in-body}
\begin{equation}\label{eq_pathloss}
P_\mathrm{L}(x,l_v)= 
   e^{-\mu_v d_v} \times \left(\frac{\lambda_v}{4\pi d_v}\right)^2 \times
   e^{-\mu_t d_t} \times \left(\frac{\lambda_t}{4\pi d_t}\right)^2  \times
   e^{-\mu_v d_s} \times \left(\frac{\lambda_s}{4\pi d_s}\right)^2,
\end{equation}
where $\mu_v$, $\mu_t$, $\mu_s$ are the molecular absorption coefficients, and $\lambda_v$, $\lambda_t$, and $\lambda_s$ are the effective wavelengths of the vessel, tissue, and skin, respectively.
The dependency with the coordinates $x$ and $l_v$ are explicitly given with the three distances as
\begin{align}
\label{eq:distances}
d_v=\frac{L_v}{\cos\varphi} \quad , \quad 
  d_t=\frac{L_t}{\cos\varphi} \quad \small{\textrm{and}} \quad 
  d_s=\frac{L_s}{\cos\varphi},    \\ 
\small{\textrm{with}} \quad   
   \cos\varphi=\sqrt{1-\left(\frac{v_{l_v}t}{l_v+L_t+L_s}\right)^2}, \nonumber 
\end{align}

\begin{wrapfigure}{R}{0.6\linewidth}
	\centering
	\includegraphics[width=\linewidth]{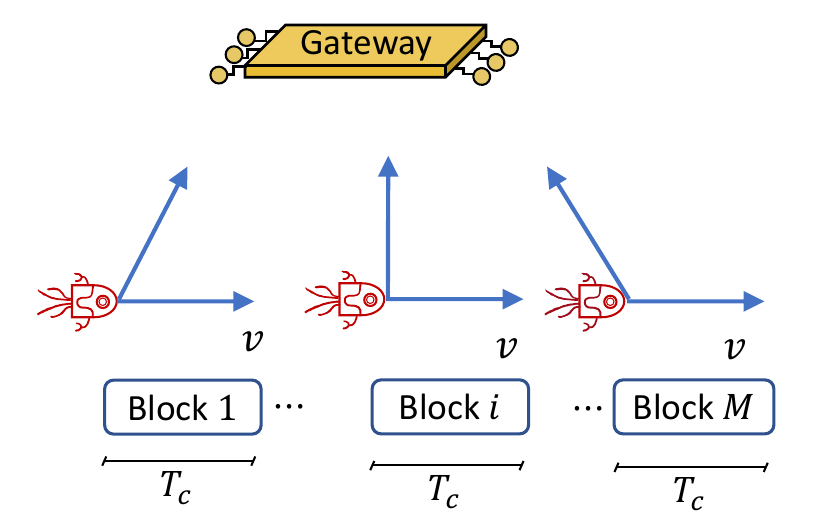}
	\caption{Transmission blocks of the length of the coherence time transmitted at different positions \cite{torres-gomez2023optimizing}.}
	\label{fig_system_model_3}
\end{wrapfigure}

The expression in \eqref{eq_pathloss} can be further reduced with the practical assumption that the nanosensor only performs emissions along the channel coherence time; see representation with $T_c$ in \Cref{fig_system_model_3}.
This way, the time dependence with \mbox{$x=v_{l_v}t$} can be omitted to particular positions along the horizontal axis, which we denote with the discrete variable $m$.
The variable $m$ also represents the sequence of information blocks in the range $[1,\, M]$, where bits are transmitted without the distortions introduced by the Doppler effect.
The variable $M$ is the maximum number of transmission blocks we can allocate in this time frame as $M=\frac{\Delta T}{T_c}=\frac{2x_\mathrm{max}}{vT_c}$, where $x_\mathrm{max}$ denotes the maximum communication range with the gateway, as illustrated in \Cref{fig_system_model_4}.
The variable $T_c$ denotes the channel coherence time and can be calculated as 
\begin{equation}\label{eq:coherence_time}
    T_c=\sqrt{\frac{9}{16\pi}}\frac{1}{\nu_\mathrm{max}},
\end{equation}
as follows from~\cite [Eq. 4.40.c]{rappaport2009wireless}, where $\nu_\mathrm{max}=v\frac{f_c}{c}$ is the maximum Doppler shift, $v$ is the transmission speed, $f_c$ is the center frequency, and $c$ is the speed of light.

Following the above description for the \ac{SNR} dependency with the random variable $l_v$ and a particular transmission block $m$, the \ac{PDF} for the \ac{SNR} can be evaluated with the Leibniz rule as in \cite[Eq. (5-5)]{papoulis2002probability}
\begin{equation}\label{eq_pdf_gamma}
    f_{\gamma_\mathrm{m}}=-f_{L_v}\frac{dl_v}{d\gamma_\mathrm{m}}\bigg|_{l_v=\gamma_\mathrm{m}^{-1}(l_v)},    
\end{equation}
which completes the description of the terahertz channel model.

%

\subsection{Evaluating the BER Performance}

We can approximate the \ac{BER} by first evaluating the probability of decoding a received symbol with errors, as indicated in \cite{carlson2002communication}.
Assuming a symbol comprises $B$-bits, the \ac{BER} is readily approximated by
\begin{equation}\label{eq_BER}
    \mathrm{BER}\approx \frac{p_\mathrm{m,e}}{B}
\end{equation}
where $p_\mathrm{m,e}$ denotes the probability of decoding with error the symbol accounting for the transmission block $m$.

The probability $p_\mathrm{m,e}$ can be evaluated with the probability that the received \ac{SNR} ($\gamma_\mathrm{m}$) becomes less than the \ac{SNR} threshold $\gamma_0$, as set by the gateway circuit.
The value for $\gamma_0$ depends on the sensitivity of the decoding circuit and the particular technology.
Analytically, $p_\mathrm{m,e}$ is formulated as 
\begin{equation}\label{eq:outage_prob}
    p_\mathrm{m,e}=P[\gamma_\mathrm{m}<\gamma_0]=\int_0^{\gamma_0}{f_{\gamma_\mathrm{m}}d\gamma_\mathrm{m}},
\end{equation}
where the integral is evaluated using \eqref{eq_pdf_gamma} as follows
\begin{equation}\label{eq_outage_prob_2}
\begin{split}
    p_\mathrm{out,m} =\int_{\gamma_{0,\mathrm{m}}^{-1}(l_v)}^{L_v}{f_{l_v}dl_v} =1-\frac{\gamma_{0,\mathrm{m}}^{-1}(l_v)}{L_v},
\end{split}
\end{equation}
and $\gamma_{0,\mathrm{m}}^{-1}(l)$ refers to the inverse relation of $\gamma_\mathrm{m}$ with $l_v$ when equating $\gamma_{0}=\gamma_\mathrm{m}$ in \Cref{eq_SNR_1}.
This inverse relation must be solved numerically for $l_v$ when $\gamma_\mathrm{m}(l_v)=\gamma_0$.

\begin{figure}
	\centering
	\includegraphics[width=0.7\columnwidth]{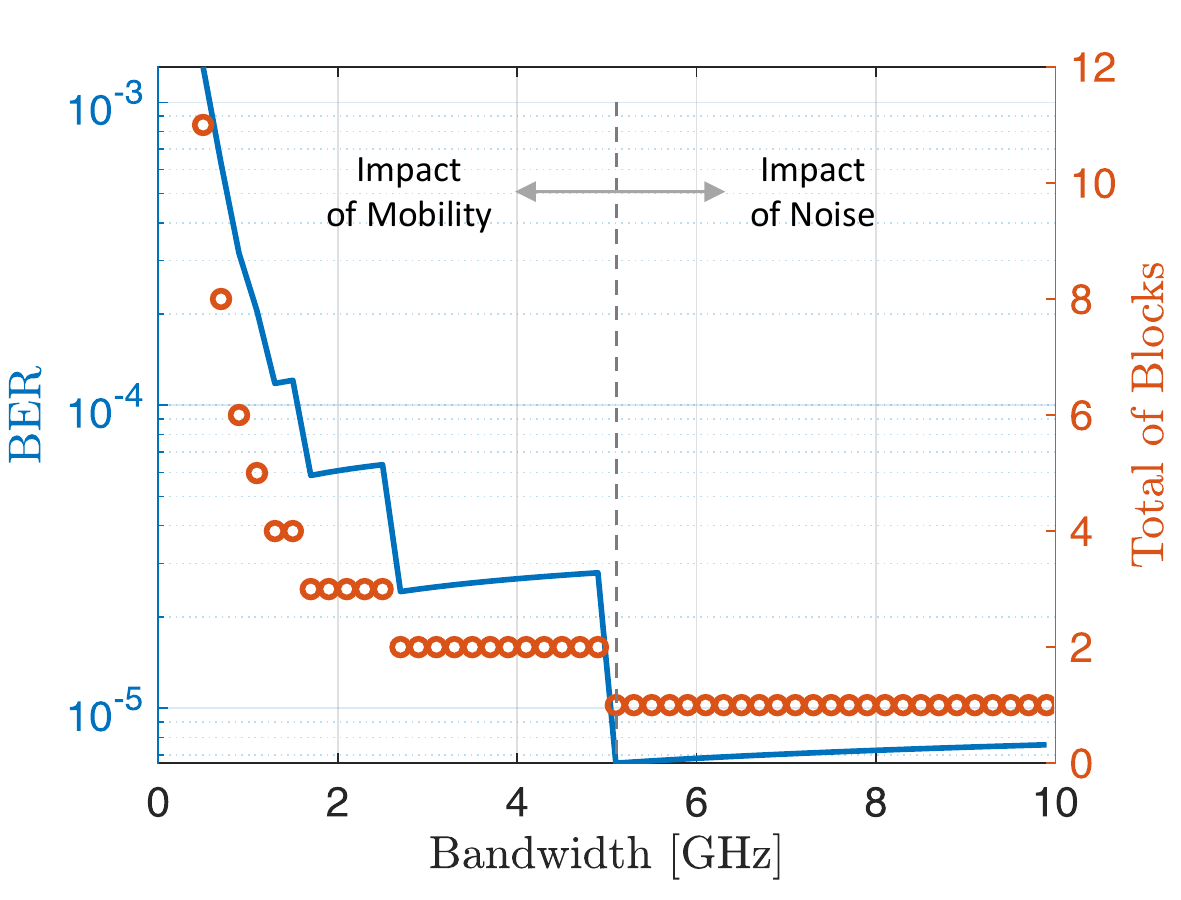}
	\caption{\small Resulting \ac{BER} and total of blocks $M$ considering the parameters in \Cref{tab_parameters_1} \cite{torres-gomez2023optimizing}.}
	\label{fig_ber_vs_bit}
\end{figure}

We illustrate the evaluation for the \ac{BER} with the plot in \Cref{fig_ber_vs_bit} and the parameters in \Cref{tab_parameters_1}.
On the horizontal axis is the transmission bandwidth, and two vertical axes reflect the \ac{BER} and the number of blocks we can allocate within the communication range with the gateway.
When considering its dependency on the bandwidth, we can identify the trade-off between noise and the impact of mobility.
Transmitting the less bandwidth increases the \ac{BER} as more symbols need to be transmitted at a position shifted from below the gateway, where the path loss is the least.
Transmitting with a larger bandwidth eventually allows transmitting all the bits in a single symbol.
However, increasing it above the minimum needed ($\SI{5}{\giga\hertz}$) also starts expanding the \ac{BER} as a result of the increased noise.
The minimum is observed at a bandwidth where the nanosensor mobility and noise impact are balanced.

This result leads us to a consequential conclusion: Transmission must be performed below the gateway and with the minimum bandwidth for transmitting a symbol with the least noise impact.
This conclusion also realizes the need to develop a transmission strategy that combines the spatial position of the nanosensor with that of the gateway.
Conceptually, the synchronization mechanism should be implemented to profit from this conclusion.
Following this finding, we elaborate on a low-complexity synchronization method in the next section.

\begin{table}
    \centering
    \caption{Simulation parameters.}
    \label{tab_parameters_1}%
    \begin{tabular}{llll}
    \toprule
    Parameter & Variable & Value & Reference \\
    \midrule
    \multicolumn{1}{p{10.32em}}{Center Frequency} & $f_c$ & $\SI{0.5}{\tera\hertz}$ &  \\
    Modulation Scheme & \acs{BPSK} &       &  \\
    \ac{SNR} threshold & $\gamma_0$ & $\SI{15}{\decibel}$ &  \\
    Packet size & $\mathrm{packet size}$ & $\SI{8}{\kilo\byte}$ &  \\
    Pulse transmission energy \textsuperscript{a}  & $E_\mathrm{Tx}$ & $\SI{5}{\milli\joule}$ & 
    \\
    \multicolumn{1}{p{10.32em}}{Blood speed in the veins} & $v$   & $\SI{0.03}{\meter\per\second}$ & \cite{guyton2015guyton} \\
    Skin thickness & $L_s$ & $\SI{86}{\micro\meter}$ & \cite{piro2016terahertz} \\
    Tissue thickness & $L_t$ & $\SI{1.44}{\milli\meter}$ & \cite{oltulu2018measurement} \\
    Vessel thickness & $L_v$ & $\SI{477}{\micro\meter}$ & \cite{fruchard2014estimation} \\
    \bottomrule
    \end{tabular}%
    \begin{tablenotes}
      \small
      \item \textsuperscript{a}\scriptsize The energy is computed as $\nicefrac{P_\mathrm{Tx}}{\mathrm{BW}}$ assuming a peak power amplitude of $P_\mathrm{Tx}=\SI{5}{\kilo\watt}$ \cite{simonjan2021in-body} and a minimum transmission rate of $\SI{1}{\giga\hertz}$.
    \end{tablenotes}
\end{table}%

%

\subsection{Practical Deployment of a Communication Scheme}

In the above section, we sketched a major conclusion: the lowest \ac{BER} is achieved when transmitting all bits with the least possible bandwidth and when the nanosensor is just below the gateway.
This strategy leads to conceiving a synchronization algorithm that triggers the communication with the location of the nanosensor.
In this section, we develop such a mechanism toward a practical deployment, which follows the contribution in \cite{torres-gomez2024low-complex}.

\begin{figure}
	\centering
	\includegraphics[width=\textwidth]{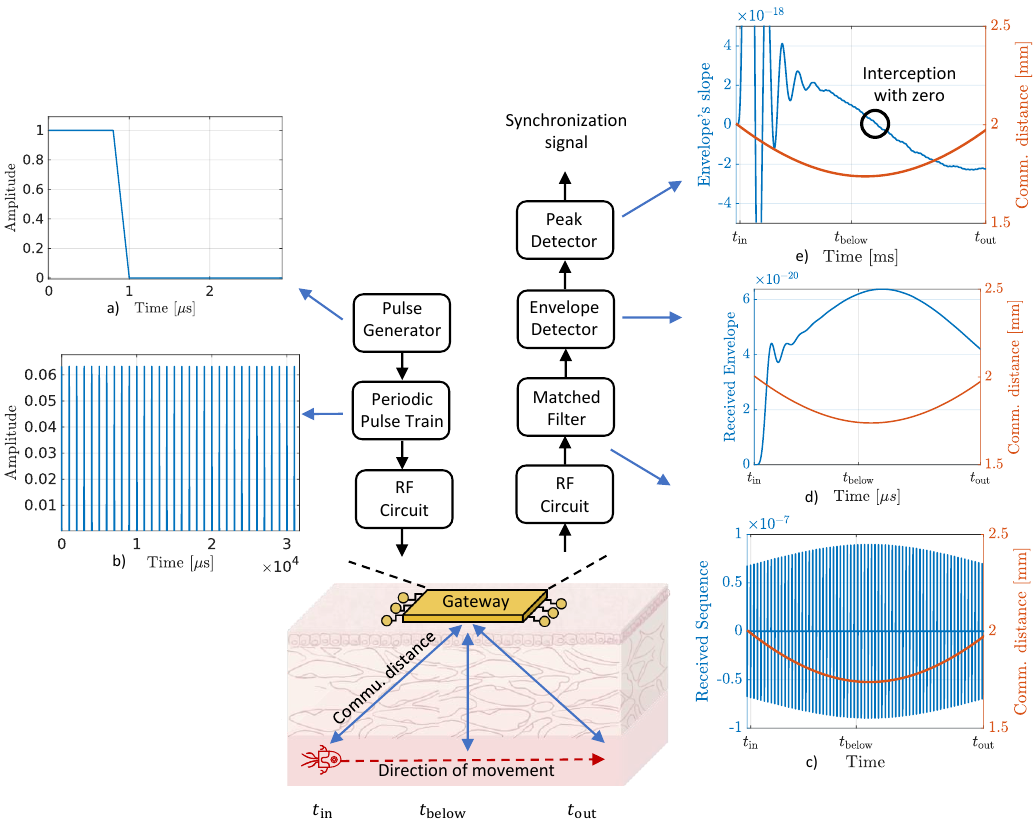}
	\caption{Overview of the synchronization components \cite{torres-gomez2024low-complex}.}
	\label{fig_receiver_scheme}
\end{figure}

Our design is deployed mainly in the gateway, as the nanosensor has low computational and transmission power.
In \Cref{fig_receiver_scheme}, we depict the emitter and receiver components at the gateway and the main signals.
Specifically, at the emitter, we have the following components:
\begin{itemize}
    \item The pulse generator block, which generates the squared signal in \Cref{fig_receiver_scheme}~a).
    \item The periodic pulse train block, which generates the pulse train signal illustrated in \Cref{fig_receiver_scheme}~b).
    \item RF circuit, which is the analog frontend and comprises the amplifier and transmitting antenna.
    Here only the amplifier is modeled.
\end{itemize}

The emitter at the gateway continuously emits a train of pulses into the tissue, which are back-scattered by the nanosensor flowing within the blood vessels.
The back-scattered signal is illustrated in \Cref{fig_receiver_scheme}~c).

The amplitude of the back-scattered signal depends on the communication distance, and its maximum will determine the precise time instant when the nanosensor is below the gateway, i.e., when the distance is the shortest.
To detect the maximum amplitude of the back-scattered signal, we implement the following blocks at the receiver side of the gateway:
\begin{wrapfigure}{R}{0.6\linewidth}
	\centering
	\includegraphics[width=.6\columnwidth]{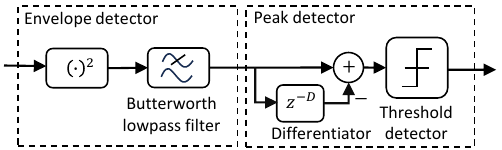}
	\caption{Synchronization scheme \cite{torres-gomez2024low-complex}.}
	\label{fig_envelope_detector}
\end{wrapfigure}
\begin{itemize}
    \item RF circuit, which is the analog front at the receiver and comprises the antenna, amplifier, and bandpass filter.
    This circuit is modeled in this work with only the receiver's sensitivity and noise power parameters.
    We avoid including an amplifier and bandpass filter for the circuit's simplicity.
    This circuit produces as the output the signal illustrated in \Cref{fig_receiver_scheme}~c).
    \item Matched filter: This filter evaluates in the output the energy of the received pulses; it follows a maximum likelihood principle for detection (see \cite{sklar2001digital}) with the desired property of producing the maximum \ac{SNR} \cite{sklar2001digital}.
    The filter develops the cross-correlation of the received signal with the expected symbol.
    We implement this filter with a \ac{FIR} structure, where its coefficients are the samples of the emitted rectangular pulse by the gateway.
    \item Envelope detector: This block follows the standard design of the analog \ac{AM} receivers (see \cite[Sec. 4.5 page 176]{carlson2002communication}) and implements a square block followed by a lowpass filter; see the representation in \Cref{fig_envelope_detector}.
    The envelope detector captures the amplitude of the oscillating signal, as illustrated in \Cref{fig_receiver_scheme}~d).
    \item Peak detector: This block implements a first-order differentiator followed by a threshold detector; see \Cref{fig_envelope_detector}. 
    The differentiator output resembles the first derivative of the amplitude signal, and the threshold detector finds its interception with zero.
    See the results of this operation in \Cref{fig_receiver_scheme}~e).
\end{itemize}

As such, the transmitter-receiver scheme in the gateway produces a synchronization-like signal at the output of the peak detector.
With this signal, the gateway triggers communication with the nanosensor, which is conveniently located at the shortest distance.

\begin{figure}
	\centering
	\includegraphics[width=.6\columnwidth]{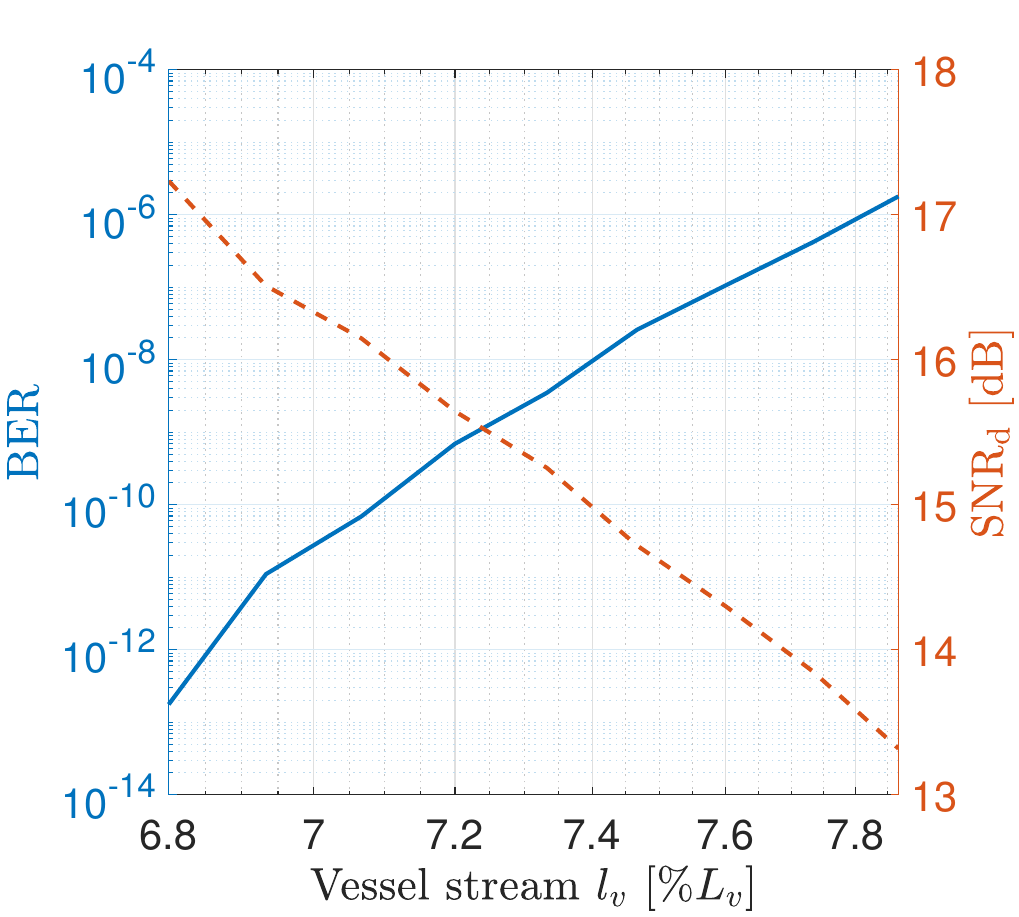}
	\caption{Comparative \ac{BER} curves versus \ac{SNR} for the ideal receiver and the synchronization scheme with zero-crossing detector when the noise power is $\sigma^2=\SI{-86.5}{\decibel}$ \cite{torres-gomez2024low-complex}.}
	\label{fig_BER_SNR_th}
\end{figure}

We illustrate the results of this implementation in \Cref{fig_BER_SNR_th}, where we plot with the vessel stream $l_v$ the resulting \ac{BER} and the postdetection \ac{SNR}, here denoted as $\mathrm{SNR}_\mathrm{d}$.
The parameters for this simulation are illustrated in \Cref{tab_parameters_2}, which follows realistic figures from hardware components and physiology parameters of the human vessels.
The results exhibit a quite favorable behavior for the \ac{BER} in the $\numrange{e-13}{e-6}$ as a result of the \ac{SNR} parameter (slightly larger than $\SI{13}{\decibel}$).

\begin{table}[b]
    \centering
    \caption{Simulation parameters}
    \label{tab_parameters_2}%
\begin{tabular}{llll}
    \toprule
    Parameter & Variable & Value & Reference \\
    \midrule
    Pulse shape & \multicolumn{1}{p{5.135em}}{Rectangular pulse} &       &  \\
    Pulse duration & $T$   & $\SI{1}{\micro\second}$ &  \\
    Pulse period & $T_p$ & $\SI{1}{\milli\second}$ &  \\
    Receiver's sensitivity &       & $\SI{0.1}{\nano\watt\per\hertz^{1/2}}$ & \cite{rogalski2019graphenebased} \\
    Noise power \textsuperscript{a}& $\sigma^2$ & $\SI{-62}{\decibel}$ & \\
    Center Frequency &  $f_c$  & $\SI{0.14}{\tera\hertz}$ & \cite{rogalski2019graphenebased} \\
    Sampling time & $T_s$ & $\SI{50}{\nano\second}$ &  \\
    \multicolumn{1}{p{10.32em}}{Blood speed in the veins} & $v$   & $\SI{0.03}{\meter\per\second}$ & \cite{guyton2015guyton} \\
    Skin thickness & $L_\mathrm{skin}$ & $\SI{76}{\micro\meter}$ & \cite{piro2016terahertz} \\
    Tissue thickness & $L_\mathrm{tissue}$ & $\SI{1}{\milli\meter}$ & \cite{oltulu2018measurement} \\
    Vessel thickness & $L_\mathrm{vessel}$ & $\SI{200}{\micro\meter}$ & \cite{fruchard2014estimation} \\
    \bottomrule
    \end{tabular}%
    \begin{tablenotes}
      \small
      \item \textsuperscript{a}\scriptsize The noise power is evaluated with the receiver's sensitivity and the transmission bandwidth; see the calculations in \cite[Sec. III]{torres-gomez2024low-complex}.
    \end{tablenotes}
\end{table}%

%

\section{Conclusions}

Advances in biomedical technologies were the primary motivation to address these two channels: ultrasonic and terahertz.
Ultrasonic waves reach the largest distance within human tissue, although they are more exposed to nanosensor mobility.
The terahertz channels are less affected but are highly impacted by the channel attenuation factor.
Although both channels achieve a low \ac{BER}, today's technological development favors terahertz technologies due to their miniaturization.
Unlike ultrasonic technologies, the terahertz design creates nanoantennas on the micrometer ($\si{\micro\meter}$) scale (see examples in \cite{torres-gomez2024implications,torres-gomez2024mobility}), placing them within a practical size range for nanosensors to navigate within the blood vessels.

%

\chapter{Evaluation of the Average PAoI Metric}
\label{sec_AoI}

\section{Fundamentals on the AoI Concept and Metrics}

The \ac{AoI} concept is constructed to evaluate the information's freshness at the monitor device regarding the status of a remote source.
Let's assume a source generates a new packet at time $t_1$, when the packet content informs about the current status of the source.
Then, the age of the source status with that packet is defined similarly to the natural age, given as the time elapsed since the creation of the packet, i.e., $(t-t_1)$, see \cite{yates2021age}.
When this packet is successfully decoded at the gateway, the monitor displays an update about the new status.
The current status is already aged with the elapsed time from the packet generation to the packet reception.
This age of the current status is precisely the definition of the \ac{AoI} concept.

\begin{figure}
	\centering
	\includegraphics[width=0.7\linewidth]{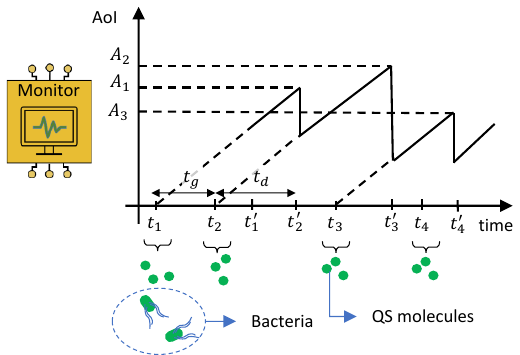}
	\caption{\ac{AoI} representation and \ac{PAoI} values as $A_1$ to $A_3$.}
	\label{fig_AoI}
\end{figure}

The illustration in \Cref{fig_AoI} clarifies further the above description.
Packets content reflects the concentration level of \ac{QS} molecules, performed with the sensing operation of the nanosensor.
These packets are generated whenever the nanosensor is within the infection location occurring at random time instants $t_1$, $t_2$, $\dots$, $t_n$.
While the nanosensor travels from the infection to the gateway location, time elapses till the corresponding $t'_1$, $t'_2$, $\dots$, $t'_n$, when the packet is successfully delivered.
The age of each packet is represented with the dashed lines in \Cref{fig_AoI}, each one starting at the generation times $t_i$.
The \ac{AoI} curve corresponds to the continuation of these dashed lines after the time of reception at $t_i'$.
Represented with a solid line, the \ac{AoI} is a sawtooth curve, giving the total of seconds elapsed since the most recent reception of the 
packet.

Based on the \ac{AoI} concept, the two most popular metrics are the average of the \ac{AoI} curve and the average of its peaks~\cite{pappas2023age}.
We follow the average of the \ac{PAoI} as it results mathematically more tractable.
The peaks of the \ac{AoI} capture the largest age the monitor experiences with the updates, represented as $A_1$, $A_2$, and $A_3$ in \Cref{fig_AoI}.
The \ac{PAoI} sequence value directly reflects the monitor's refresh period.
That is, the peaks are the periodicity of the monitor update, and the average \ac{PAoI} is the average update period of fresh information at the monitor.

The amplitude of the \ac{AoI} peaks is also related to the packet's generation time ($T_g$) and the time delay ($T_d$), based on the geometric properties of the triangles.
Looking at \Cref{fig_AoI}, the triangle $t_1$, $t'_2$, $A_1$ proves to be isosceles due to the rectangular angle between sides $\overline{t_1t'_2}$ and $\overline{t'_2A_1}$ and the~$\SI{45}{\degree}$ of the hypotenuse $\overline{t_1A_1}$ with the side $\overline{t_1t'_2}$. 
As a result, the peak amplitude $A_1$ amounts to $(t_d+t_g)$; the same result follows for the other peaks.\footnote{Here the noncapital notation as $t_d$ refers to a realization of the random variable $T_d$, and similar follows for the generation time $t_g$.}
Accordingly, the relation $\mathbf{E}[A_i]=\mathbf{E}[T_d]+\mathbf{E}[T_g]$ holds, where $T_d$ and $T_g$ are the random variables for the generation time and the delay.

Furthermore, some packets carried by the nanosensors may not be delivered
due to decoding errors in the communication interface with the gateway.
In that case, the height of the peaks in the \ac{AoI} increases as the monitor must wait for the next packet to arrive.
Denoting with $p_\mathrm{loss}$ the probability for the packet loss, a closed-form expression for the average \ac{PAoI} is given by
\begin{equation}\label{eq_PAoI}
    \Delta^{(p)}=\frac{1}{1-p_\mathrm{loss}}\mathbf{E}[T_g]+\mathbf{E}[T_d],
\end{equation}
as derived from similar principles in \cite[Sec. 3.3]{torres-gomez2022age}.

%

\section{Evaluating the Average PAoI Metric}
\label{sec_average_PAoI}

The expression in \eqref{eq_PAoI} for the average \ac{PAoI} is thoroughly evaluated with the three parameters $p_\mathrm{loss}$, $\mathbf{E}[T_g]$, and $\mathbf{E}[T_d]$.
The probability $p_\mathrm{loss}$ is already defined by the communication interfaces introduced in \Cref{sec_comm}; it amounts to the \ac{BER} if we consider that updates are delivered binary, or to the \ac{PER} if the updates are delivered in the form of a packet.
The random variables $T_g$ and $T_d$ are still to be determined from the mobility model of the nanosensor following \Cref{sec_mobility}.

\begin{wrapfigure}{R}{0.5\linewidth}
	\centering
	\includegraphics[width=\linewidth]{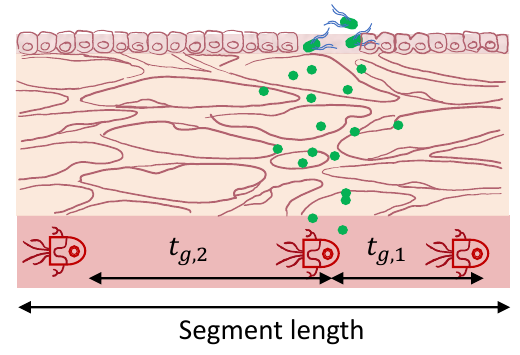}
	\caption{Visualizing the generation time period as nanosensors flow through the target~\cite{torres-gomez2023age}.}
	\label{fig_generation_time}
\end{wrapfigure}

Looking at the generation time $T_g$, this variable evaluates the periodicity that nanosensors pass through the infection location; see examples of two particular realizations with $t_{g,1}$ and $t_{g,2}$ in \Cref{fig_generation_time}.
We can consider two pragmatic assumptions to characterize $T_g$: i)~the sampling of the \ac{QS} molecules by the nanosensor is an independent process; i.e., the reading of one nanosensor is not influencing the reading of the next one, and ii) this sampling process occurs at a constant average rate, i.e., the average number of nanosensors passing through the infection location is constant with time.
Following these assumptions, then the random variable $T_g$ will follow the exponential distribution as given by (see \cite[Sec. 5.5]{ross2010first})
\begin{equation}
f_{T_g}(t_g)=\left\{
    \begin{matrix}
    \lambda_i e^{-\lambda_i t_g}, & t_g\geq 0,\\
    0\, & t_g< 0,
    \end{matrix}\right.
\end{equation}
where the average rate $\lambda_i$ refers to the average number of nanosensors passing through the infection site in vessel segment $i$ over time.
This calculation can be completed with the Markov model in \Cref{sec_mobility} and the laminar profile of the bloodstream as follows:
\begin{itemize}
    \item The number of nanosensors in the vessel segment (corresponding to the infection location) can be calculated with the stationary probability in \Cref{eq_stationary} as $N_i=\nu_i\cdot N_s$, where the index $i$ refers to the vessel segment and $N_s$ is the number of nanosensors flowing in the cardiovascular system (as follows from the Markov model).
    \item The traveling time for a single nanosensor in the vessel segment $i$ can be approximated with the average speed of the blood ($\overline{v}_i$) and the length of the vessel ($L_i$) as $T_i=\dfrac{L_i}{\overline{v}_i}$ (as follows from the laminar profile of the bloodstream).
    \item As a result, the average number of nanosensors in the unit of time is readily given by $\lambda_i=\frac{N_i}{T_i}$.
\end{itemize}
Following the above reasoning, the expected value of the generation time yields
\begin{equation}\label{eq_generation_time}
    \mathrm{E}[T_g]=\frac{1}{\lambda_i}=\frac{T_i}{N_i}.
\end{equation}

Contextualizing this formula to the physiology of the cardiovascular system, let's consider the Tibialis posterior in the legs, where injuries are common.
In this vessel segment, the blood speed in the capillaries is approximately $\SI{0.1}{\meter\per\second}$ \cite{guyton2015guyton}, and the vessel length is a maximum $\SI{2}{\centi\meter}$ around \cite{salgado2009fibula}.
Thereby, a single nanosensor takes $T_v\approx\SI{0.2}{\second}$ to flow through.
Besides, the probability of finding a single nanosensor in the Tibialis posterior amounts to $p=0.005$; see \Cref{fig_results_nanosensors_distribution}.
As a result, assuming there are a total of \num{e3} flowing nanosensors in the cardiovascular system, approximately  \mbox{$N_v\approx p\times 10^3=5$} nanosensors will be flowing through the Tibialis posterior.
Using these approximations, the expected average for the generation time will result in \mbox{$\mathrm{E}[T_g]\approx\frac{\SI{0.2}{\second}}{5}=\SI{40}{\milli\second}$}.
A similar procedure applies to evaluate the average generation time for other tissues in the body.

The second random variable, $T_d$, refers to the time it takes a nanosensor to travel from the infection location to the gateway.
We evaluate the expected value for $T_d$ fetching each different circuit in the cardiovascular system as a realization of $T_d$.
A circuit is given by a closed loop of the human vessels that follows the bloodstream.
Following the physiology of the cardiovascular system, all the closed loops pass by the heart and a given capillary set.
For instance, this is the case of the circuit right ventricle-lungs-left ventricle-capillary radialis, and similarly for the other capillaries.
As such, the human body has as many closed circuits as capillaries.
In our representation of the cardiovascular system in \Cref{fig_Markov}, there are $15$ different circuits in total.

\begin{figure}
	\centering
	\includegraphics[width=\linewidth]{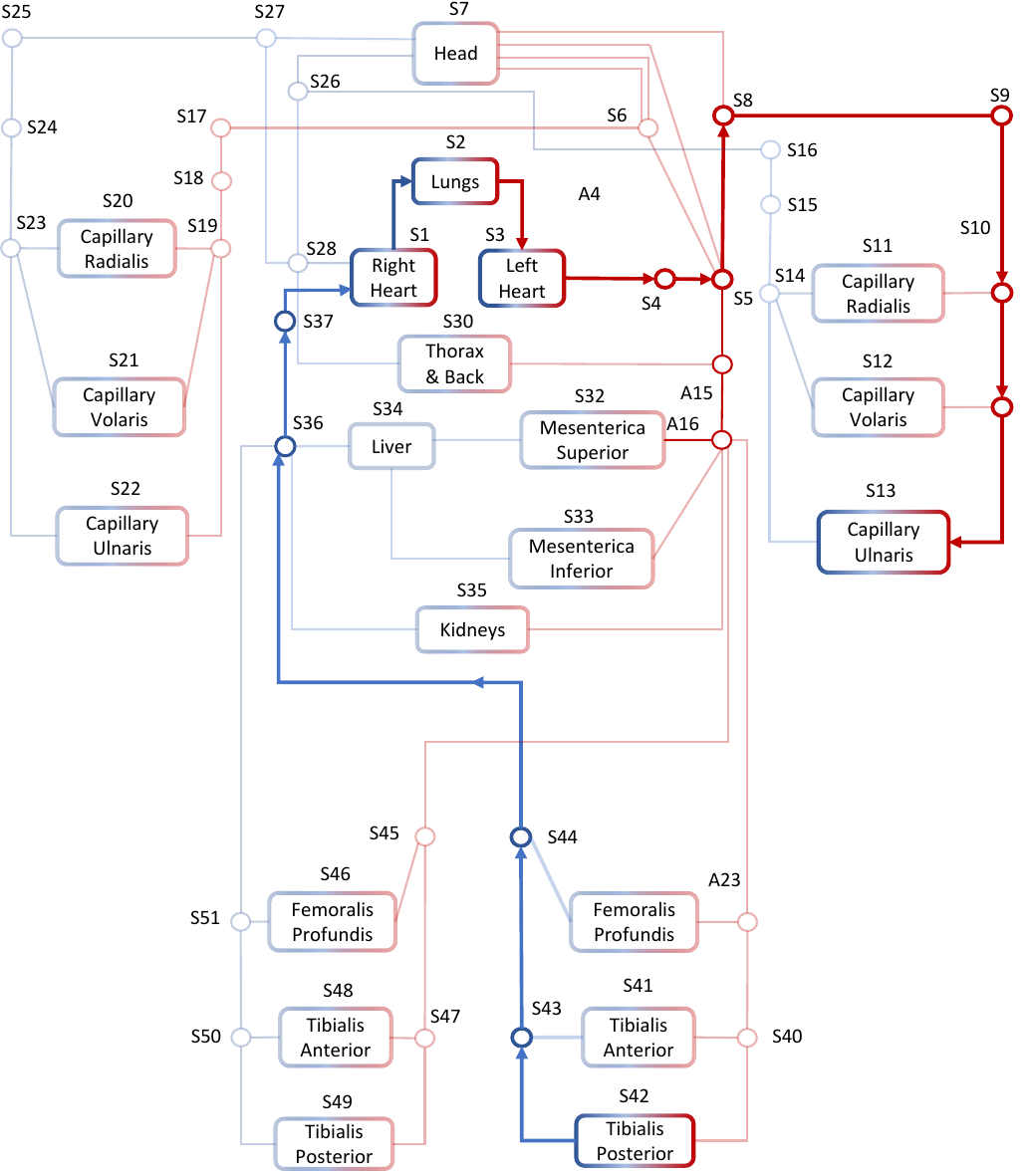}
	\caption{Highligted path in the Markov model between the infection location (in the Tibialis Posterior) and the gateway (at the Capillary Ulnaris).}
	\label{fig_markov_results}
\end{figure}

We denote the traveling time on a circuit as $T_c$, and it occurs with a probability $p_c$, where the index $c$ denotes the circuit.
Thereby, $E[T_d]$ equates to $\sum_c T_c\cdot p_c$ --with the allowance of the informal notation for simplicity.
Let's illustrate this calculation without loss of generality when the infection is in the Tibialis Posterior and the monitor is on the Ulnaris; similar reasoning can be applied to other location pairs.
As we highlight in \Cref{fig_markov_results}, there is a direct path from the infection location to the gateway.
This path will evaluate a delay, here denoted as $T_{S42\to S13}$ and occurring with the probability that the nanosensor follows all these vessel bifurcations, here denoted as $p_{S42\to S13}$.

The probability $p_{S42\to S13}$ can be evaluated with the transition probability of the Markov model.
In this case, the nanosensor travels with probability one to the left heart, and from there with probability \mbox{$p_{S42\to S13}=p_{S5\to S8}\cdot p_{S8\to S9}\cdot p_{S9\to S10}\cdot p_{S10\to S13}$} to the gateway location at the Ulnaris (referred to the wrist in the left arm); see the highlighted path in \Cref{fig_markov_results}.

However, at this point, it's worth mentioning that once the nanosensor finds its place in the left heart, the bloodstream might steer it to follow a different path, as a circuit $c'$, instead of continuing to the gateway or the infection.
In that case, the nanosensor completes a round along a different circuit in the time $T_{c'}$ before reaching the left heart again.
As a result, the average traveling time from the infection to the gateway location must evaluate all these options, here calculated as 
\begin{equation} 
\begin{split}
    \label{eq_delay}
    \mathrm{E}[T_d] & = T_{S42\to S13}\cdot p_{S42\to S13} + \sum_{\substack{c\\c\neq S42\\c\neq S13}}\sum_{k=1}^{\infty}{(k\cdot T_c+T_{S42\to S13})\cdot p_c^k},
    \end{split}
\end{equation}
where the first term refers to the direct travel between the infection and the gateway circuit, and the second term accounts for the sum in $c$ for other circuits than the infection and the gateway.
The sum in $k$ for the possibly infinite number of completed loops along the circuits.

\begin{figure}
	\centering
	\includegraphics[width=0.8\linewidth]{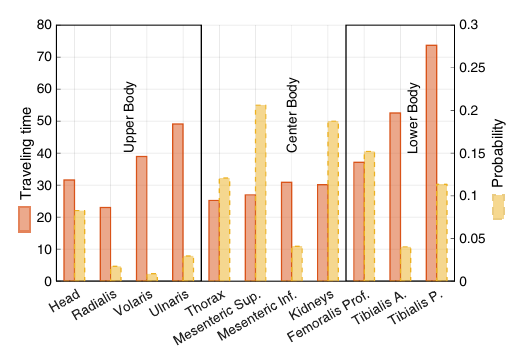}
	\caption{Average traveling time along the various circuits in the cardiovascular system and probabilities per closed loop in the human cardiovascular system \cite{torres-gomez2023age}.}
	\label{fig_probabilities}
\end{figure}

\Cref{eq_delay} can be further reduced using the MacLaurin series expansion formulas $\sum_{k=1}^\infty{p^k}=\frac{p}{1-p}$ and $\sum_{k=1}^\infty{kp^k}=\frac{p}{(1-p)^2}$ for the sum in $k$, yielding the closed form expression
\begin{equation}
\begin{split}
\label{eq_delay_2}
    \mathrm{E}[T_d] =
    T_{S42\to S13}\cdot p_{S42\to S13}+ \sum_{\substack{c\\c\neq S42\\c\neq S13}}T_{c}\frac{p_c}{(1-p_c)^2}+T_{S42\to S13}\frac{p_c}{1-p_c}.
    \end{split}
\end{equation}
To evaluate this variable, we illustrate in \Cref{fig_probabilities} the probabilities $p_c$ and the traveling time on each circuit.
The probabilities are evaluated from the Markov model; the traveling time per circuit is evaluated with the \ac{BVS} simulator.\footnote{We provide open access to the code at \url{https://github.com/jorge-torresgomez/BVS_data} and also document the corresponding \ac{BVS}'s dataset following to \textcite{gebru2021datasheets}.}
Finally, with Eqs. \eqref{eq_delay_2} and \eqref{eq_generation_time} we can evaluate the average \ac{PAoI} having the probability $p_\mathrm{loss}$ as a parameter.

\begin{figure}
 \centering
 \subfloat[Monitor at the heart.]
 {\includegraphics[width=0.32\linewidth]{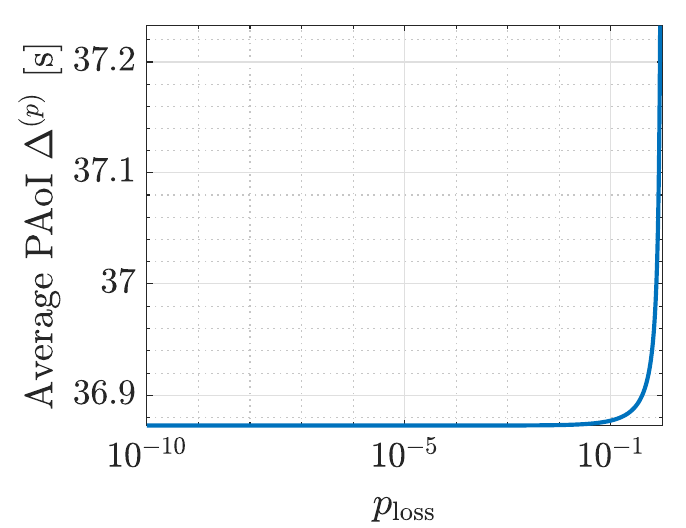}
 \label{fig_LIME_slope_a}}
 \subfloat[Monitor at the left wrist.]
 {\includegraphics[width=0.3\linewidth]{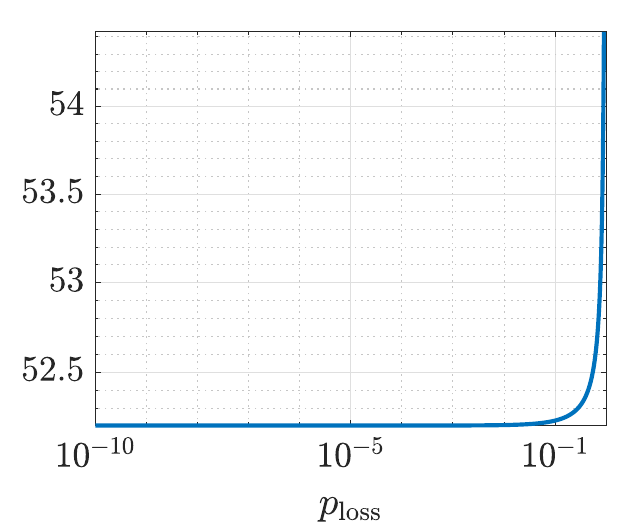}
 \label{fig_res_AoI_b}}
 \subfloat[Monitor at the femoralis.]
 {\includegraphics[width=0.3\linewidth]{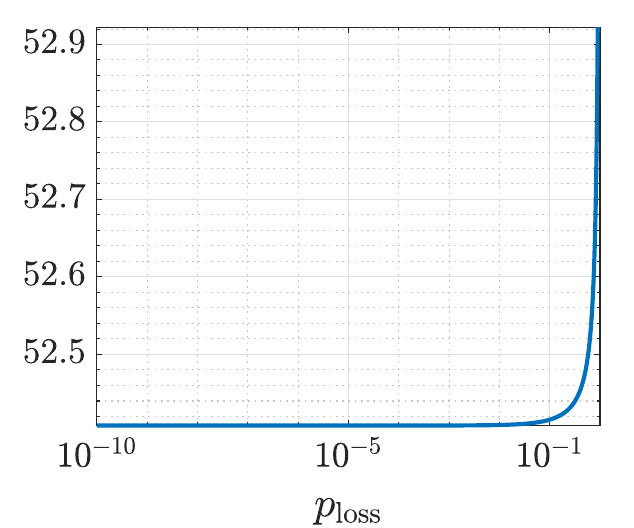}
 \label{fig_res_AoI_c}}
 \caption{Average \ac{PAoI} when the gateway is located at the heart, the left wrist, and the Femoralis Profundis.
 The infection location is at the Tibialis Posterior \cite{torres-gomez2023age}.}
 \label{fig_res_AoI}
\end{figure}

We illustrate in \Cref{fig_res_AoI} the resulting metric from three different monitor placements, assuming a total of \num{e3} nanosensors within the cardiovascular system.
This figure draws these main conclusions:
\begin{itemize}
    \item The highest information freshness is achieved at the heart (update period slightly below the $\SI{37}{\second}$ when delivery losses are low), as the heart is always a center node to all the circuits in the cardiovascular system.
    \item Locating the monitor at the wrist or the femoralis produces an almost similar update period.
    Although the femoral is farther from the heart than the wrist, the blood speed to the femoral is higher due to the main arteries connecting it (the thorax and femoral aortas).
    This way, the largest distance to the femoral is compensated with a higher blood speed.
    \item Surprisingly, the resulting average \ac{PAoI} metric is independent of the \ac{PER} in a wide range.
    Whenever the \ac{PER} is less than around \num{e-1}, the average \ac{PAoI} remains the same.
    This results from the shorter generation time (on a millisecond scale) compared to the delivery time (in units of seconds).   
    Thereby, missing some packets will extend the waiting time to receive the next one, but not considerably.
\end{itemize}

%

\section{Conclusions}

We used the average \ac{PAoI} metric to unify the randomness of sampling and information delivery on a single formulation.
This metric reflects the monitor update period, jointly accounting for sampling time and delivery of updates.
Information freshness depends on the frequency of sample generation, which occurs when the nanosensor meets the infection location.
The refresh period depends on the capability to deliver the sample, which relates to the physical process of information traveling and delivering.
Merging generation and traveling time realize the freshness of updates in the network. 
Surprisingly, our results exhibit little dependency on the update period with the quality of delivery; see \Cref{fig_res_AoI}.

%

\chapter{Conclusions and Outlook}
\label{sec_future}

\section{Conclusions}

Exploring how much time matters in \ac{IoBNT} networks, this thesis looks at assessing \ac{AoI} metrics and provides orders-of-magnitude values for information freshness.
Research progress on \ac{IoBNT} is quite mature today, allowing us to examine from the system-level consideration \acl{QoS} metrics related to their monitoring capabilities. 
As a factor metric, we assess the information's freshness with the \ac{AoI} concept.
The average \ac{PAoI} metric evaluates the time elapsed between samples displayed on the monitor and their generation time.
Freshness indicates a real-time display of events, and this metric directly assesses how well the monitor represents the live occurrence of events at a remote source.
Within the \ac{IoBNT} scenario evaluated in this thesis, this metric illustrates the network capability to portray the real-time evolution of biomarkers within the human body. 

\textbf{Relevancy of the concept:}
We also remark on the unmistakable definition of the \ac{AoI} concept towards information freshness compared to other standard metrics like delay or throughput.
The \ac{IoBNT} network can be designed with a low delay, however it can still display outdated samples.
The delay only concerns the channel and eludes the periodicity of transmissions.
Thus, the monitor will mismatch the source's real-time developments if the source is limited to low transmission rates (low throughput).
Freshness, interpreted in terms of delay and throughput metrics, realizes the current concept for the \ac{AoI}. 

The \ac{AoI} encapsulates in a compact formulation the various random references within the network: generation of samples, carrying, and delivery.
Within the \ac{IoBNT} network we portray in this thesis, the generation time will refer to the nanosensor sampling process of biomarkers, their traveling through the network of the cardiovascular system, and their delivery to the external monitor.
As such, this concept places together all processes from sampling to displaying.

This formulation is functional in analyzing these variables' interrelation and balancing network resources.
The information freshness can be enhanced by increasing the sampling rate or reducing the network delays.
However, a solution must be adjusted to the particularities of the \ac{IoBNT} network --increasing the sampling rate includes increasing the number of nanosensors (which can be harmful to the human body) and decreasing the traveling time can be achieved with a better placement of gateways (which can be limited by communication constraints through the human body).
As such, this concept opens up pathways for further architectural design that is adequate to the environment of the human body.

\textbf{Assessing the time sensitivity: }
Results indicate that the monitor is outdated by tens of seconds regarding events occurring at the largest distances from body parts.
Within the \ac{IoBNT} architecture in this thesis, real-time monitoring is effectively conveyed whenever the event occurring within the human body is comparatively slower, up to hundreds of seconds.
This is the case of bacterial infections at the tissue level, where their evolution occurs on a minute-time scale.
However, at the cellular level, where processes occur in less than tens of seconds, more adequate architectures are needed to shorten the outdated periods.

\textbf{Paths forward: }
The development of this concept is also a platform for integrating multiple research domains, including human physiology (needed to characterize the information flow within the human body) and network layers (from the physical to the application layer).
The development requires elaborating on the integration of communication schemes within the dynamics of human physiology.
We described integrating the nanosensors mobility through the human vessels in \Cref{sec_comm} and their communication components in \Cref{sec_mobility}; both evaluate the average \ac{PAoI} metric, as described in \Cref{sec_average_PAoI}.
A different \ac{IoBNT} architecture, for instance, with the placement of various gateways or the inclusion of fusion nodes, will require more developments (theoretical and simulators) toward their integration.

\section{Future Research Directions}

We also identify two potential future research directions from a system-level perspective:

\begin{enumerate}
    \item \textbf{Optimizing performance:}
    Thus far, we provide within this thesis a theoretical framework to evaluate the information freshness in \ac{IoBNT} networks.
    From this point, a noticeable research direction is formulating optimization problems to maximize information freshness while balancing network resources.
    The problem will be more concerned with the architecture and constraints.
    For instance, the placement of gateways or the deployment of fusion nodes are network components that require formulation within the \ac{AoI} framework; see preliminary results in \cite{pal2024age}.
    Furthermore, pertinent constraints must be formulated, such as the number of nanosensors in the system and their lifetime.
    Too many would harm the human body, and in practice, they operate for a limited period due to battery constraints or malfunctioning.
    \item \textbf{Anticipating the events:} The results illustrated in the above chapter indicate that information freshness is on the order of tens of seconds along the most extended trajectories; see \Cref{fig_res_AoI}.
    This magnitude order allows monitoring events at the tissue scale involving multiple cells.
    For instance, in an aerated culture, bacteria duplicate their population number every $\SIrange{20}{25}{\min}$, see \cite[page 279]{alberts2013essential}.
    Information freshness in the seconds' order magnitude would be sufficient in this time range.
    However, at the cell level, live processes occur much faster: the ribosomes in the cell take around $\SI{20}{\second}$ to several minutes to synthesize proteins (\cite[page 232]{alberts2013essential}) and the protein components (like the amino acids) are assembled in the units of milliseconds (\cite[page 229]{alberts2013essential}).
    Monitoring such a process would require a more adequate \ac{IoBNT} architecture to reduce the timing or to include anticipatory mechanisms on the monitor side.
    For instance, with a model of the underlying chemical process, a monitor could forecast developments and get ahead of the received outdated samples.
    Models and the accuracy of anticipatory mechanisms are open challenges for keeping the monitoring fresh.
\end{enumerate}

%

\part{Published Papers}
\pagenumbering{gobble}

\section[Paper A]{Paper A: ``Nanosensor Location Estimation in the Human Circulatory System using Machine Learning''}

Published as:

\begin{itemize}
    \item Jorge Torres Gómez, Anke Kuestner, Jennifer Simonjan, Bige Deniz Unluturk and Falko Dressler, ``Nanosensor Location Estimation in the Human Circulatory System using Machine Learning,'' IEEE Transactions on Nanotechnology, vol. 21, pp. 663–673, October 2022.
\end{itemize}

Author’s Contribution (J. Torres Gómez):
\begin{itemize}
    \item Conceived and formulated the research concept and objectives.
    \item  Designed the machine learning methods for nanosensor location estimation and performed the core simulations.
    \item Analyzed the experimental results and interpreted their significance in the context of \ac{IoBNT}.
    \item Coordinated writing efforts: drafted the main sections and integrated co-author feedback.
\end{itemize}

Co-authors assisted with domain-specific expertise on machine learning frameworks, provided partial data processing scripts, and contributed sections of the final manuscript edits.


%




\section[Paper B]{Paper B: ``Electric Circuit Representation of the Human Circulatory System to Estimate the Position of Nanosensors in Vessels''}

Published as 
\begin{itemize}
    \item Jorge Torres Gómez, Jorge Luis González Rios and Falko Dressler, ``Electric Circuit Representation of the Human Circulatory System to Estimate the Position of Nanosensors in Vessels,'' Elsevier Nano Communication Networks, vol. 40, pp. 100499, July 2024.
\end{itemize}

Author’s Contribution (J. Torres Gómez):

\begin{itemize}
    \item Originated the idea of applying an electric-circuit-based model (“Frankenstein simulator”) for blood-flow dynamics.
    \item Developed the circuit modeling approach and performed code implementation for simulations.
    \item Led the data collection, parameter-tuning experiments, and result analysis.
    \item Prepared the initial manuscript draft and incorporated co-authors’ comments.
\end{itemize}

Co-authors provided electrical modeling advice, proofread the circuit representations, and contributed improvements to the discussion and conclusion sections.



%

\section[Paper C]{Paper C: ``Low-Complex Synchronization Method for Intra-Body Links in the Terahertz Band''}

Published as 
\begin{itemize}
    \item Jorge Torres Gómez, Jennifer Simonjan and Falko Dressler, ``Low-Complex Synchronization Method for Intra-Body Links in the Terahertz Band,'' IEEE Journal on Selected Areas in Communications, vol. 42 (8), pp. 1967–1977, August 2024.
\end{itemize}

Author’s Contribution (J. Torres Gómez):

\begin{itemize}
    \item Proposed the synchronization framework tailored for terahertz intra-body communication.
    \item Derived the theoretical analysis and outlined performance metrics.
    \item Programmed the simulation setup and conducted the experiments that validated the synchronization method.
    \item Wrote the manuscript draft, with sections on methodology and results, and responded to reviewer comments.
\end{itemize}

Co-authors refined aspects of the synchronization protocol, provided additional simulations for cross-checking, and critically reviewed all manuscript versions.



%

\section[Paper D]{Paper D: ``Optimizing Terahertz Communication Between Nanosensors in the Human Cardiovascular System and External Gateways''}

Published as 
\begin{itemize}
    \item Jorge Torres Gómez, Jennifer Simonjan, Josep Miquel Jornet and Falko Dressler, "Optimizing Terahertz Communication Between Nanosensors in the Human Cardiovascular System and External Gateways," IEEE Communications Letters, vol. 27 (9), pp. 2318–2322, September 2023.
\end{itemize}

Author’s Contribution (J. Torres Gómez):

\begin{itemize}
    \item Conceived the optimization problem and formulated it in terms of terahertz channel constraints and physiological requirements.
    \item Implemented numerical algorithms, executed the optimization runs, and interpreted their outcomes.
    \item Edited most of the sections in the manuscript and integrated co-author feedback on advanced terahertz channel modeling.

\end{itemize}

Co-authors contributed domain expertise in terahertz band channel modeling, assisted with advanced theoretical derivations, and reviewed the complete manuscript.



%

\section[Paper E]{Paper E: ``Age of Information-based Performance of Ultrasonic Communication Channels for Nanosensor-to-Gateway Communication''}

Published as 
\begin{itemize}
    \item Jorge Torres Gómez, Joana Angjo and Falko Dressler, "Age of Information-based Performance of Ultrasonic Communication Channels for Nanosensor-to-Gateway Communication," IEEE Transactions on Molecular, Biological and Multi-Scale Communications, vol. 9 (2), pp. 112–123, June 2023.
\end{itemize}

Author’s Contribution (J. Torres Gómez):

\begin{itemize}
    \item Created the framework linking ultrasonic channel modeling and Age of Information (AoI) analysis.
    \item Performed the derivation of performance metrics and coded the simulation environment.
    \item Analyzed the data, authored the main sections of the paper (introduction, methodology, results), and integrated co-author suggestions.
\end{itemize}

Co-authors provided insights on ultrasonic wave propagation specifics and the \ac{AoI} modeling, improved the final paper organization, and revised the text for clarity.



%

\section[Paper F]{Paper F: ``Age of Information in Molecular Communication Channels''}

Published as 
\begin{itemize}
    \item Jorge Torres Gómez, Ketki Pitke, Lukas Stratmann and Falko Dressler, "Age of Information in Molecular Communication Channels," Elsevier Digital Signal Processing, Special Issue on Molecular Communication, vol. 124, pp. 103108, May 2022.
\end{itemize}

Author’s Contribution (J. Torres Gómez):

\begin{itemize}
    \item Proposed the original idea of applying \ac{AoI} principles to molecular communication systems.
    \item Led the theoretical analysis of generation, delivery, and signal impairment for channel modeling.
    \item Designed, implemented and documented the simulation pipeline.
    \item Drafted the manuscript and integrated co-authors’ text contributions.
\end{itemize}

Co-authors supplemented conceptual discussions on molecular signaling processes, assisted with drafting a portion of the theoretical background, and refined the final text presentation.



%

\cleardoublepage

\listofabbreviations
\clearpage

\listoffigures
\clearpage

\listoftables
\clearpage

\printbibliography

\end{document}